# Large exchange bias after zero-field cooling from an unmagnetized state

B. M. Wang,[1] Y. Liu,[1,*] P. Ren,[2] B. Xia,[2] K. B. Ruan,[2] J. B. Yi,[3] J. Ding,[3] X. G. Li,[4] and L. Wang[2,†]

[1]*School of Mechanical and Aerospace Engineering, Nanyang Technological University, 639798, Singapore*

[2]*Division of Physics and Applied Physics, School of Physical and Mathematical Sciences, Nanyang Technological University, 637371, Singapore*

[3]*Department of Materials Science and Engineering, National University of Singapore, 119260, Singapore*

[4]*Hefei National Laboratory for Physical Sciences at Microscale and Department of Physics, University of Science and Technology of China, Hefei 230026, China*

Electronic mail: *MLiuY@ntu.edu.sg, †WangLan@ntu.edu.sg

**Abstract:**

Exchange bias (EB) is usually observed in systems with interface between different magnetic phases after *field cooling*. Here we report an unusual phenomenon in which a *large* EB can be observed in Ni-Mn-In bulk alloys after *zero-field cooling from an unmagnetized state*. We propose this is related to the newly formed interface between different magnetic phases during the initial magnetization process. The magnetic unidirectional anisotropy, which is the origin of EB effect, can be created *isothermally* below the blocking temperature.

When a system consisting of ferromagnetic (FM)-antiferromagnetic (AFM) [1], FM-spin glass (SG) [2], AFM-ferrimagnetic (FI) [3], and FM-FI [4] interface is cooled with field through the Néel temperature ($T_N$) of the AFM or glass temperature ($T_{SG}$) of the SG, exchange bias (EB) is induced showing a shift of hysteresis loop [$M$(H)] along the magnetic field axis. Since its discovery by Meiklejohn and Bean in 1956 [1], EB has been extensively studied during the past fifty years, partly because of its applications in ultrahigh-density magnetic recording, giant magnetoresistance and spin valve devices [5, 6]. The EB effect is attributed to an FM unidirectional anisotropy formed at the interface between different magnetic phases [5]. Generally, the process of field cooling (FC) from higher temperature is used to obtain FM unidirectional anisotropy in different EB systems [1-4]. The FM unidirectional anisotropy can also be realized by depositing the AFM layer onto a saturated FM layer [5], by ion irradiation in an external magnetic field [7], by zero-field cooling (ZFC) with remnant magnetization [8, 9]. In a word, the FM unidirectional anisotropy in these EB systems is formed by *reconfiguring* the FM spins at the interface between different magnetic phases. Here, we named the previous EB generally observed after FC as the conventional EB (CEB). Furthermore, Saha *et al*. [10] argued that a small spontaneous EB observed after ZFC without remnant magnetization, which has been ignored or attributed to the experimental artifact, can be explained theoretically in an otherwise isotropic EB system. The CEB effect after FC has also been observed in NiMn-based Heusler bulk alloys, such as NiMnSn [11], NiMnSb [12], and NiMnIn [13], coexisting of AFM and FM phases. In this Letter, we report a *large* EB effect (the maximum EB field is about 1300 Oe at 10 K) after *ZFC from an unmagnetized state* in Ni-Mn-In bulk alloys. Namely, a large FM unidirectional anisotropy can be produced *isothermally*, which has never been reported to date and cannot be expected in the CEB systems [14].

The details of sample preparations and experiment measurements for $Ni_{50}Mn_{50-x}In_x$ (NiMnIn$x$, $x=$ 11, 12, 13, 14 and 15) alloys are illustrated in the supplementary information [14]. Two measurement processes can be used to obtain a closed *M*(H) loop after ZFC (only consider |+*H*| = |-*H*|):

(1) P type, 0 → (+*H*) → 0 → (-*H*) → 0 → (+*H*),

(2) N type, 0 → (-*H*) → 0 → (+*H*) → 0 → (-*H*).

The first 0 → (+*H*)/(-*H*) curve is called as an initial magnetization curve. Generally, these two kinds of measurement will obtain the same loop except for the initial magnetization curve. Thus, only one of them has been used to obtain *M*(H) loop in the previous studies. However, they will give the different results in the present study.

Figure 1(a) shows the temperature dependence of magnetization [*M*(T)] of NiMnIn13 ($T_N \sim$ 410 K) measured under *H* =10 Oe after ZFC and FC. The ZFC curve exhibits a peak at $T_p$ = 53 K and an irreversibility between ZFC and FC curves occurring at $T_f \sim$ 150 K, which is similar to that of NiCoMnSn [15]. The magnetic state of NiMnIn13 at low temperatures is superparamagnetic (SPM) domains embedded in AFM matrix as in NiCoMnSn. The SPM domains are collectively frozen forming a superspin glass (SSG) state at lower temperatures [15]. The *M*(H) curve at 300 K is a straight line without any SPM/FM feature, which indicates that the $T_c$ is at lower temperature [inset of Fig. 1(a)]. To further confirm this SSG state, we measured ac susceptibility at various frequencies ($f$s) with an ac magnetic field of 2.5 Oe after ZFC from 300 K. Figure 1(b) shows the temperature dependence of the real part of ac susceptibility. The $T_p$ increases with increasing frequency, which can be fitted to a critical power law for SSG [16],

$$\tau = 1/2\pi f = \tau^* (T_p/T_g-1)^{-z\upsilon} \qquad (1)$$

where $\tau^*$ is the relaxation time of individual particle moment, $T_g$ is the static glass temperature and $z\upsilon$ is the dynamic critical exponent. Our data can be fitted well by Eq. (1) with $\tau^* \approx 10^8$ s, $z\upsilon \approx 9.7$ and $T_g \approx 52$ K [inset of Fig. 1(b)]. These values are close to those reported for SSG ($\tau^* \approx 10^8$ s and $z\upsilon \approx 10.2$) [17]. Furthermore, the memory effect of SSG state has also been observed in NiMnIn13 [14].

The CEB effect after FC is observed in all NiMnIn$x$ ($x$= 11, 12, 13, 14 and 15) bulk alloys [14]. In this Letter, we investigate the $M$(H) loops at 10 K *after ZFC from an unmagnetized state* in these alloys. The unmagnetized initial state at 10 K in these alloys can be obtained easily if they are zero-field cooled from 300 K due to their $T_c$s are lower than 300 K [14]. Figure 1(c) shows the P type $M$(H) of NiMnIn13 at 10 K after ZFC from 300 K with maximum measurement field $|H_m^{max}|$ (=|+$H$|=|-$H$|) = 40 kOe. The dashed line shows the initial magnetization curve, which lies outside the major hysteresis loop. The magnetization at the starting point of the initial magnetization curve ($H$ = 0) is zero, indicating that the initial state at 10 K is an unmagnetized state [14]. It is worth noting that the ZFC $M$(H) loop shows a *large* shift along the magnetic field axis, which has never been observed in any previous CEB systems. The equal magnetization values in the highest positive and negative magnetic fields indicate the shifted loop is not a nonsymmetrical minor hysteresis loop [14]. We also measured the N type $M$(H) loops with opposite direction of the initial magnetization field at 10 K after ZFC [Fig. 1(d)] [14], which shift to the positive magnetic field axis showing a centrally symmetric image of the P type $M$(H) loops. This result cannot be expected from the effect of remanent field of superconductor magnet/remanent magnetization of the samples, in which the shift direction of $M$(H) loop is independent of the direction of the initial magnetization field. Furthermore, both the EB field ($H_{EB}$) and coercivity ($H_c$) after ZFC can be larger than those after FC [Fig. 1(c)] [14], which indicates that the EB after ZFC in the present case is not a spontaneous EB [10]. The $H_{EB}$ and $H_c$ are defined as $H_{EB}$ = - ($H_L$+$H_R$)/2 and $H_c$ = - ($H_L$ - $H_R$)/2, respectively, where $H_L$ and $H_R$ are the left and right coercive fields. To further confirm this phenomenon after ZFC, we measured the temperature dependence of $H_{EB}$ and $H_c$ for $|H_m^{max}|$ = 40 kOe and the training effect at 10 K for several selected $|H_m^{max}|$s [14], which are similar to those in the CEB systems obtained after FC [5, 18]. The key difference is that EB in NiMnIn13 can be observed after *ZFC from an unmagnetized state*. Namely, FM unidirectional anisotropy, usually obtained by FC from higher temperature, can be induced *isothermally* during the initial magnetization process.

To investigate its origin, we consider the evolution of the initial magnetic state of NiMnIn13 after ZFC under external magnetic field as shown in Fig. 2. It is a simplified schematic diagram with SPM domains embedded in an AFM single domain (AFM anisotropy axis is parallel to the direction of applied magnetic field). The applied magnetic field aligns all the SPM domains along the direction of external field. When the Zeeman energy of AFM spins ($J_{ZE}$, which is proportional to the magnitude of magnetic field) near the interface is larger than the coupling energy of SPM-AFM at the interface ($J_{int}$, constant) and their anisotropy energy (constant), the applied field will align these AFM spins along the direction of external field [19]. Therefore, the SPM domains will grow in size. However, the enlarged SPM domains are at *metastable state* and the coupling interface of SPM-AFM remains unchanged at this stage [see the dashed white circles in Fig. 2]. After removal of external magnetic field, they will shrink and return to their initial sizes due to the AFM anisotropy energy.

The growth of SPM domain size will decrease the inter-domain distance, thus increases the interaction between SPM domains. This is similar to the process of increasing the concentration of SPM nanoparticles in the conventional SSG systems [16]. When the interaction between SPM domains reaches the critical value, the coupling of SPM domains will become superferromagnetic (SFM) exchange through tunnelling superexchange [16]. The difference between SFM and conventional FM is that the atomic spins in the conventional FM are replaced by the superspins of SPM domains. *The formation of SFM exchange may change the internal interaction of each enlarged SPM domain (metastable at SSG state) such that they become stable as shown in Fig. 2.* [While in the case of SPM *nanoparticles* embedded in AFM matrix with a chemical interface (different materials) [6], the SPM nanoparticles cannot grow to form larger stable particles at the expense of AFM matrix]. As a result, a *new stable* SFM-AFM interface with unidirectional moment of SFM is formed and will pin the SFM superspins below the blocking temperature ($T_B$), which is similar to an FM-AFM interface with unidirectional FM spins formed after FC in the CEB systems. The difference is that in the present case the SFM-AFM interface is induced isothermally by external magnetic field. While in the CEB systems it is usually reconfigured under FC. According to this model, the moment of SPM domains increases with increasing size under external magnetic field.

We have only considered AFM domains with anisotropy axis parallel to external magnetic field in this model. For AFM domains with anisotropy axis non-parallel to external magnetic field, there is an angle between the direction of the initial magnetization field and the anisotropy axis. This configuration can still result in EB effect, which is similar to the EB effect in the CEB systems with different angles between the direction of the cooling field and the AFM anisotropy axis [20]. Based on the above analyses, we believe that a SFM unidirectional anisotropy, which is similar to an FM unidirectional anisotropy, can be formed during the initial magnetization process in the present alloys.

In order to confirm this model, we further measured the $M(H)$ loops with various magnitudes of the initial magnetization fields (different $|H_m^{max}|$s) at 10 K after ZFC from 300 K [14]. Figure 3(a) shows $H_{EB}$ and $H_c$ as a function of $|H_m^{max}|$. There is a critical $|H_m^{max}|$ ($H^{crit}$) = 30 kOe, at which $H_{EB}$ reaches the maximum value and $H_R$ remains almost constant at higher $|H_m^{max}|$ [Fig. 3(b)]. The maximum $H_{EB}$ means the formation of the maximum FM unidirectional anisotropy [5]. Thus, the meaning of $H^{crit}$ is that at which SSG state completely transforms to SFM state, producing maximum SFM unidirectional anisotropy. The decrease of the $H_{EB}$ at higher $|H_m^{max}|$ is only due to the decrease of the $H_L$ [Fig. 3(b)], which may originate from the change of bulk AFM spin structure under large applied magnetic field [14]. The bulk AFM spin structure has been shown to play a crucial role in EB effect in thin film system [21].

When $|H_m^{max}| < H^{crit}$, only part of the SSG state transforms to the SFM state during the initial magnetization process. For SSG state, there is a remnant magnetization and $H_c$ in $M(H)$ loops due to irreversible switching of a collective state [17]. The $H_c$ (both $H_L$ and $H_R$) increases with increasing $|H_m^{max}|$ (a series of minor hysteresis loops). However, the number of SPM domains at SSG state will decrease with increasing $|H_m^{max}|$ due to more SSG state transforming to SFM state, which generates more new interfaces with SFM unidirectional anisotropy. Thus, the $H_{EB}$ increases with increasing $|H_m^{max}|$ at this stage, leading to the increase of coercive field in one direction ($H_L$) and the decrease in the other ($H_R$). The final coercive fields are attributed to a combined effect of SPM domains in SSG and SFM states. Due to the opposite $|H_m^{max}|$ dependence for these two effects, the $H_R$ reaches maximum at a field smaller than $H^{crit}$ [Fig. 3(b)]. For $H_L$, both effects have the same $|H_m^{max}|$ dependences, resulting in a continuous increase of $H_L$ with $|H_m^{max}|$. As a result, the field, at which $H_c$ reaches maximum, is smaller than the $H^{crit}$ of $H_{EB}$. Further supports to the model shown in Fig. 2 are provided in the supplementary information including anomalous remanent magnetization dependence of EB effect, isothermal tuning of EB after ZFC from an unmagnetized state, and strong cooling field dependence of CEB effect in NiMnIn13 [14].

Finally, we have further verified the model by changing the size of the initial domains, which is crucial to the formation of the SFM unidirectional anisotropy [14]. If the size of the initial domains is larger than the critical value, SFM or FM domains will form and no EB effect will appear after ZFC [13]. Figure 4(a) shows the $M$(H) curves of NiMnIn$x$ alloys at 10 K. The saturation magnetization of NiMnIn$x$ increases with increasing In content [Fig 4(b)], which is consistent with the previous results [22]. The $T_c$s of these alloys are lower than 300 K and the $T_N$ decreases continuously with increasing In content [14]. The saturation magnetization of NiMnIn$x$ at 10 K is very small compared with that of the stoichiometric compound $Ni_{50}Mn_{25}In_{25}$ (80 emu/g, pure FM state at low temperatures) [22]. The decrease of saturation magnetization in the off-stoichiometric alloys is due to the excess of Mn atoms occupying a number of In sites, which produces AFM coupling [22]. The SFM (may include some SPM or FM domains) volume fraction increases from ~1 % in NiMnIn11 to ~ 35 % in NiMnIn15 at 10 K [Fig. 4(b)]. Thus, the average domain size in $x = 14$ alloy is larger than that of NiMnIn13 at the initial state. The larger size of SPM domain makes the $H^{crit} = 15$ kOe, at which all of the SSG state transforms to SFM state, being smaller than that of NiMnIn13 [Fig. 4(c)]. Furthermore, the SFM volume fraction in $x = 14$ alloy is about 22 % (the total volume fraction of SPM/SFM in the initial state is less than this value), which is close to the threshold concentration for percolation in three dimensional system (~16 %) [23]. The SFM domains no longer separate from each other in AFM matrix at larger volume fraction resulting in the formation of FM domains at $x > 14$. For $x = 15$ alloy, there is no SPM domains at the initial state and the $M$(H) loops after ZFC shows double-shifted behavior with no EB effect, which is similar to the results of NiMnIn16 [13, 14]. For $x = 11$ and 12, the continuous increase of $H_{EB}$ with $|H_m^{max}|$ up to 80 kOe is due to the smaller size of SPM domain, which is similar to results of NiMnIn13 for $|H_m^{max}| < H^{crit}$. Large $H_c$ has also been observed for $x=11$ and 12 and the $H_c$ of NiMnIn12 shows tendency to maximum value at higher $|H_m^{max}|$s prior to the maximum of $H_{EB}$ [Fig. 4(d)]. All of these results are consistent to the discussions in NiMnIn13 within the model as shown in Fig. 2.

In summary, we have observed a large EB effect after ZFC from an unmagnetized state in Ni-Mn-In bulk alloys, exhibiting the same relationship of the temperature dependence of $H_{EB}$ and $H_c$, and the training effect as in the CEB systems after FC. Such phenomenon is attributed to a SFM unidirectional anisotropy formed during the initial magnetization process. These results will open a new direction to realize EB effect.

Authors are indebted to C. Leighton for stimulating discussions. Support for this work came from Singapore National Research Foundation (RCA-08/018) and MOE Tier 2 (T207B1217).

**Figure captions:**

FIG. 1. (color online) (a) $M$(T) curves measured under $H$ =10 Oe after ZFC and FC. The inset shows the $M$(H) curve at 300 K. (b) Temperature dependence of the real part of the ac susceptibility measured at frequencies $f$ = 0.1, 1, 10, 100, 1000 Hz with ac magnetic field of 2.5 Oe after ZFC from 300 K. The inset shows the plot of $\log_{10} f$ vs $\log_{10} (T_p/T_g-1)$ (open circles) and the best fit to Eq. (1) (solid line). (c) and (d) $M$(H) loops of NiMnIn13 at 10 K with $|H_m^{max}|$ = 40 kOe after ZFC and FC ($H$ = 40 kOe) from 300 K. The dashed lines show the initial magnetization curves. The insets show the larger scale at the low field.

FIG. 2. (color online) Simplified schematic diagrams of the evolution of the SPM domains embedded in an AFM single domain (AFM anisotropy axis is parallel to the direction of the applied magnetic field) under external magnetic field at temperature below $T_B$. The initial magnetic state after ZFC is a SSG state. The white arrows represent the superspin direction of SPM domains. The dashed white circles show the coupling interfaces of SPM-AFM. The dashed blue lines represent that the coupling of SPM domains is a glassy coupling, while the solid blue lines represent that the coupling of SPM domains is a SFM exchange.

FIG. 3. (color online) (a) $H_{EB}$ (left panel) and $H_c$ (right panel) as a function of $|H_m^{max}|$ in NiMnIn13 at 10 K after ZFC. (b) The left ($H_L$) and right ($H_R$) coercive fields as a function of $|H_\text{m}^\text{max}|$. Inset shows the definition of $H_\text{L}$ and $H_\text{R}$ in a $M(H)$ loop. The dot lines show the position of $H^\text{crit}$ = 30 kOe.

FIG. 4. (color online) (a) Field dependence of magnetization of NiMnIn$x$ ($x$ = 11, 12, 13, 14, and 15) at 10 K. (b) Saturation magnetization ($M^\text{sat}$) and SFM volume fraction as a function of In content at 10 K. (c) and (d) $|H_\text{m}^\text{max}|$ dependence of $H_\text{EB}$ and $H_\text{c}$ at 10 K for $x$ = 11, 12 (right panel), and 14, 15 (left panel).

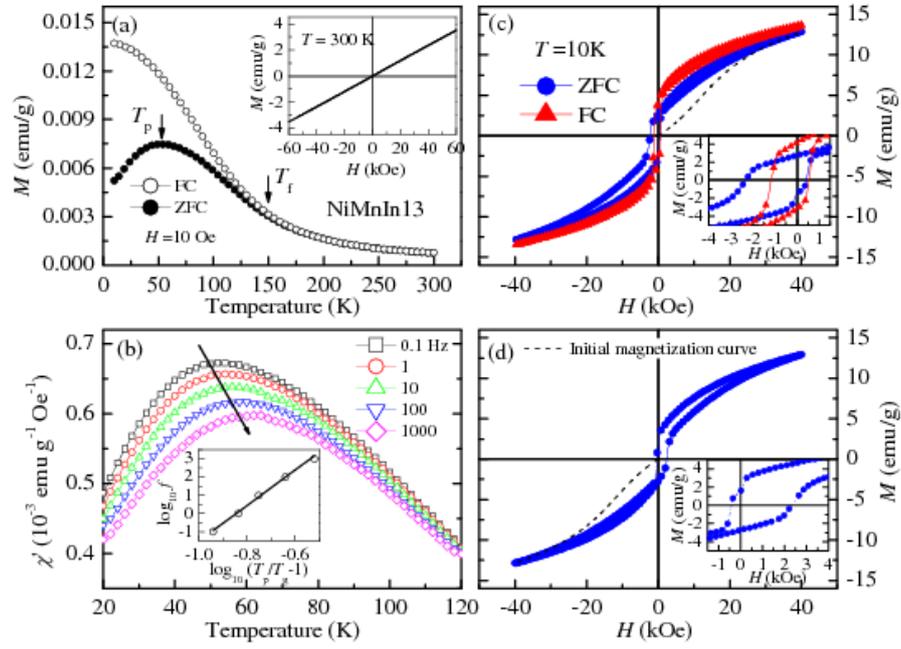

FIG. 1. B. M. Wang *et al.*

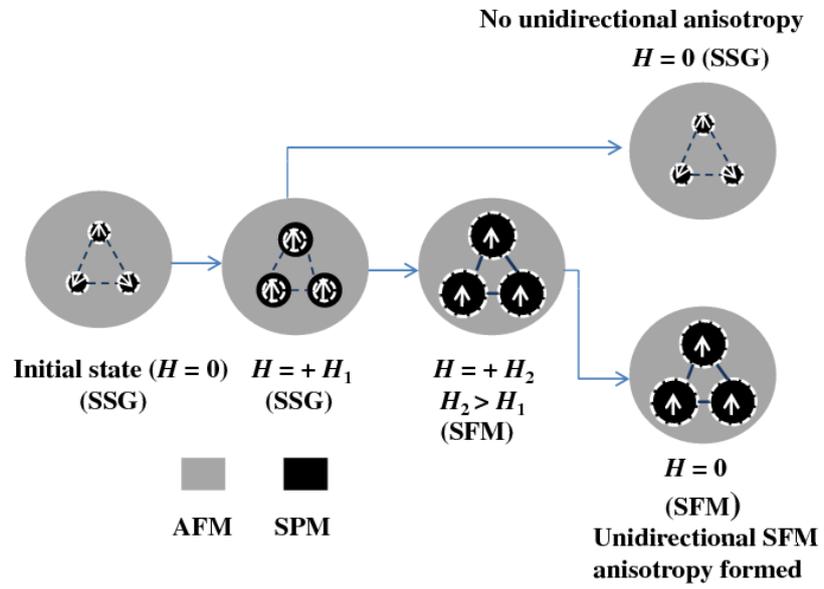

FIG. 2. B. M. Wang *et al.*

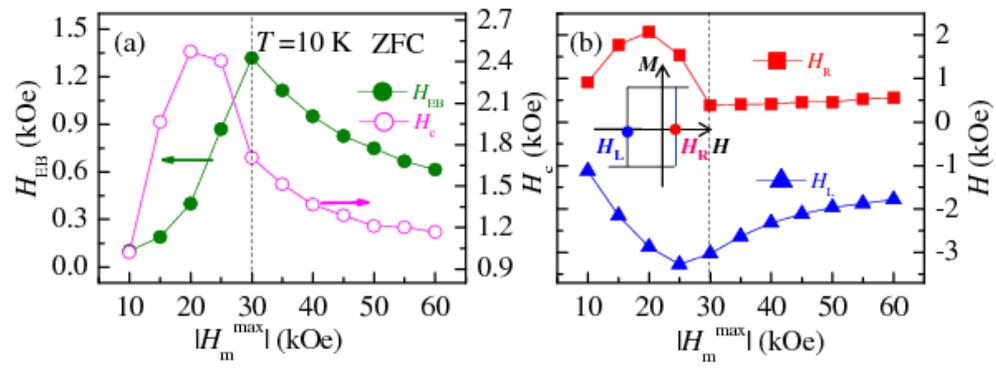

FIG. 3. B. M. Wang *et al.*

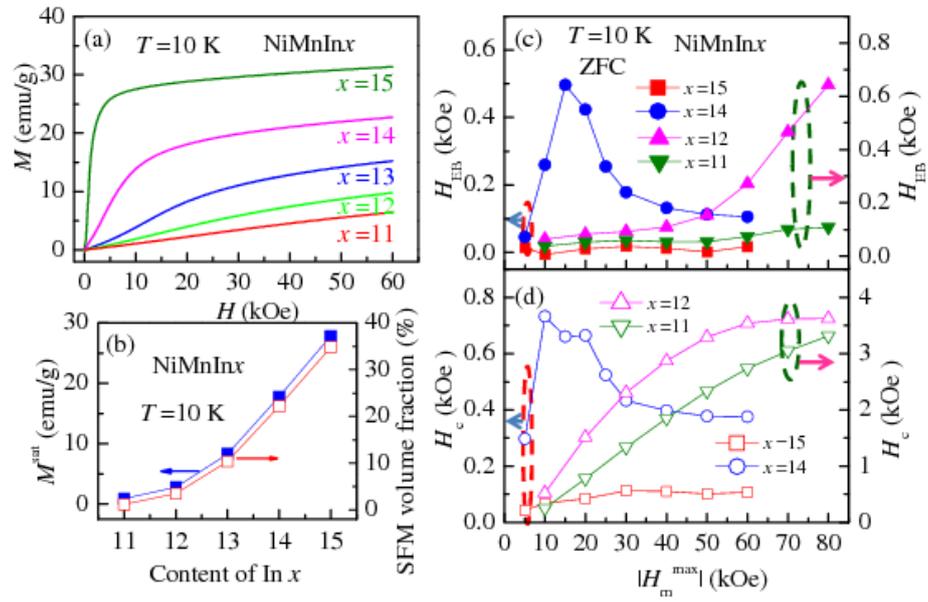

FIG. 4. B. M. Wang *et al.*

[16] S. Bedanta, and W. Kleemann, J. Phys. D: Appl. Phys. **42**, 013001 (2009).

[17] S. Sahoo, O. Petracic, W. Kleemann, S. Stappert, G. Dumpich, P. Nordblad, S. Cardoso, and P. P. Freitas, Appl. Phys. Lett. **82**, 4116 (2003).

[18] M. K. Chan, J. S. Parker, P. A. Crowell, and C. Leighton, Phys. Rev. B **77**, 014420 (2008).

[19] C. Leighton, J. Nogués, H. Suhl, and I. K. Schuller, Phys. Rev. B **60**, 12837 (1999).

[20] J. Nogués, T. J. Moran, D. Lederman, I. K. Schuller, and K. V. Rao, Phys. Rev. B **59**, 6984 (1999).

[21] R. Morales, Z. P. Li, J. Olamit, K. Liu, J. M. Alameda, and I. K. Schuller, Phys. Rev. Lett. **102**, 097201 (2009).

[22] T. Krenke, M. Acet, E. F. Wassermann, X. Moya, L. Mañosa, and A. Planes, Phys. Rev. B **73**, 174413 (2006).

[23] H. Scher, and R. Zallen, J. Chem. Phys. **53**, 3759 (1970).

# Supplementary information

## Large exchange bias after zero-field cooling from an unmagnetized state

B. M. Wang,[1] Y. Liu,[1,*] P. Ren,[2] B. Xia,[2] K. B. Ruan,[2] J. B. Yi,[3] J. Ding,[3] X. G. Li,[4] and L. Wang[2,†]

[1]*School of Mechanical and Aerospace Engineering, Nanyang Technological University, 639798, Singapore*

[2]*Division of Physics and Applied Physics, School of Physical and Mathematical Sciences, Nanyang Technological University, 637371, Singapore*

[3]*Department of Materials Science and Engineering, National University of Singapore, 119260, Singapore*

[4]*Hefei National Laboratory for Physical Sciences at Microscale and Department of Physics, University of Science and Technology of China, Hefei 230026, China*

Electronic mail: *MLiuY@ntu.edu.sg, †WangLan@ntu.edu.sg

## S1. Experiment methods

The polycrystalline $Ni_{50}Mn_{50-x}In_x$ (NiMnIn$x$) ($x=$ 11, 12, 13, 14 and 15) alloys were prepared with Ni, Mn and In of 99.9% purity using arc melting technique under argon atmosphere. The samples were remelted several times and subsequently annealed at 1000 ºC under high vacuum for 24 h and finally slowly cooled to room temperature to ensure homogeneity. The phase purity and crystal structure were determined by *X*-ray diffraction and the compositions were determined by energy dispersive *X*-ray analysis. The antiferromagnetic (AFM) transition temperature ($T_N$) was determined by differential scanning calorimetry. The dc magnetic properties were measured using a physical properties measurement system (PPMS, Quantum Design) with a vibrating sample magnetometer (VSM) module. The ac susceptibility was measured using a superconducting quantum interface device (SQUID, Quantum Design). Before each ZFC, the superconductor magnet was demagnetized by oscillating fields at 300 K (in PPMS) or heated up to its superconductor transition temperature (in SQUID, magnet reset) to remove the pinned magnetic flux.

## S2. Magnetic hysteresis loop after ZFC from an unmagnetized state

**a. In the conventional EB (CEB) systems with a fixed interface after ZFC (assuming an FM-AFM interface)**

The interface between FM and AFM is fixed in the CEB systems after fabrication. When the system is zero-field cooled from an unmagnetized FM state to $T < T_B$, the FM domains are in random orientations and the net magnetization is zero. Here, we simplify it as two FM domains with opposite directions parallel to the direction of the magnetic field [Fig. S1(1)]. The AFM spins next to the FM align ferromagnetically due to the interaction at the interface assuming FM interaction between FM and AFM (it can be AFM interaction [1]). The total net magnetization is zero in the initial state in the system [Fig. S1(1)]. When a positive field (right direction) is applied, the FM spins with left direction start to rotate. If AFM anisotropy is sufficiently large, the AFM spins remain unchanged. The rotated FM spins are exerted a microscopic torque, due to FM interaction between FM and AFM at the interface, to keep them in their original position (left direction) [Fig. S1(2)]. After removal of the positive field, there is a remanent magnetization due to magnetic interaction among FM domains [Fig. S1(3)]. When the negative field is applied and removed, the same process occurs for FM spins with right direction. Although the *local* FM-AFM interaction is unidirectional in the CEB systems after ZFC from an unmagnetized state, the hysteresis loop representing a collective effect of whole FM-AFM interaction is centrally symmetric (no EB effect). If the AFM anisotropy is not very large, the AFM spins near the interface will rotate when the FM spins are rotated by the magnetic field [2, 3]. However, the rotated AFM spins will return to their initial positions due to AFM anisotropy energy after removal of the external magnetic field. Then, they will still exert a microscopic torque on the FM spins due to FM interaction between FM and AFM at the interface, which is similar to the case of AFM spins remain unchanged during FM rotation resulting in a symmetric hysteresis loop (no EB effect).

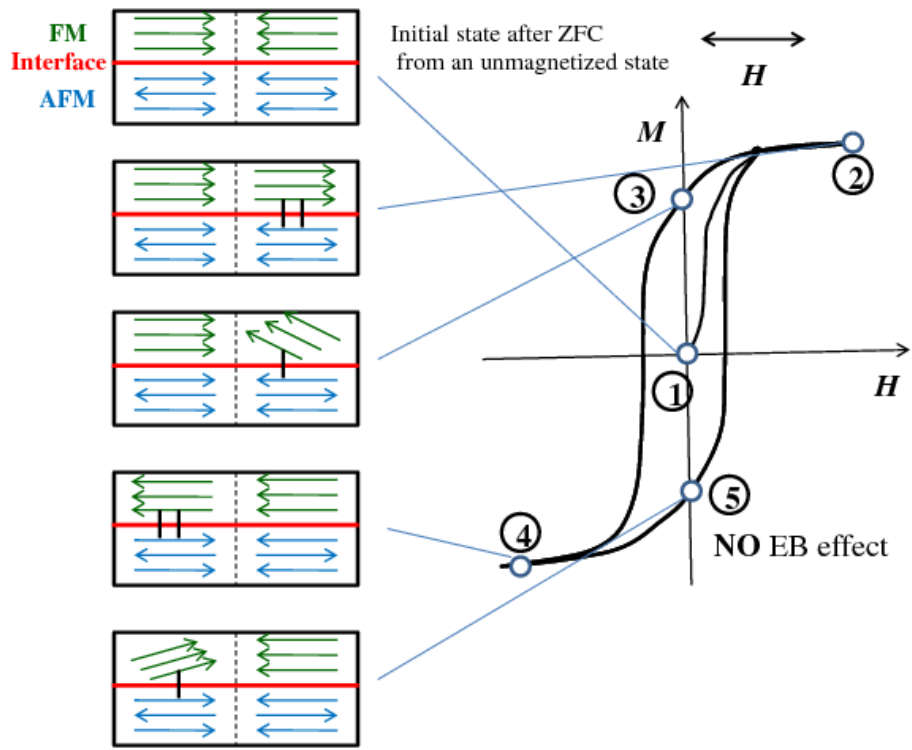

FIG. S1. Hysteresis loop of the CEB systems (FM-AFM) with a fixed interface after ZFC from an unmagnetized state (right) and schematic diagram of the spin configuration at different stages (left). Note that the spin configurations are a simple cartoon to illustrate the effect of the coupling.

**b. In the case with a variable interface after ZFC**

In order to be understood well, the same simple cartoon has been used to describe the spin configurations of the Fig. 3 in the main text. There are two assumptions: (1) FM domains grow in size at the expense of AFM domain during the initial magnetization process and the enlarged FM domains (both size and direction) can be preserved after removal of the field. Namely, the FM-AFM interface can be changed isothermally; (2) For the subsequently sweeping field, the newly formed FM-AFM interface remains unchanged. Let us see what will happen for hysteresis loop measurements in this case. The initial state after ZFC from an unmagnetized state is the same as Fig. S1(1) [Fig. S2(1)]. When a positive field is applied (initial magnetization process), the FM spins with left direction start to rotate. The AFM spins near the interface will rotate following FM rotation due to the FM interaction between FM and AFM at the interface [2, 3]. If the part of AFM domain with rotated spins become a new part of FM domain and the AFM spins at the new interface is the same as FM spins (Assumption 1), there is no microscopic torque exert on the rotated FM spins (domain on the right side) at this stage [Fig. S2(2)], which is different from the case in the CEB system [Fig. S1(2)]. When the field is reversed, the FM spins (both left and right domains) start to rotate. If the newly formed FM-AFM interface remains unchanged at this stage (Assumption 2), the AFM spins at the interface exert a microscopic torque to the rotated FM spins to keep them in their original position [Fig. S2(4)]. The FM spins in this case have only one stable configuration, which is similar to the case in the CEB systems after FC [4]. That is, if these two assumptions are satisfied, the FM unidirectional anisotropy can be formed isothermally after ZFC from an unmagnetized state.

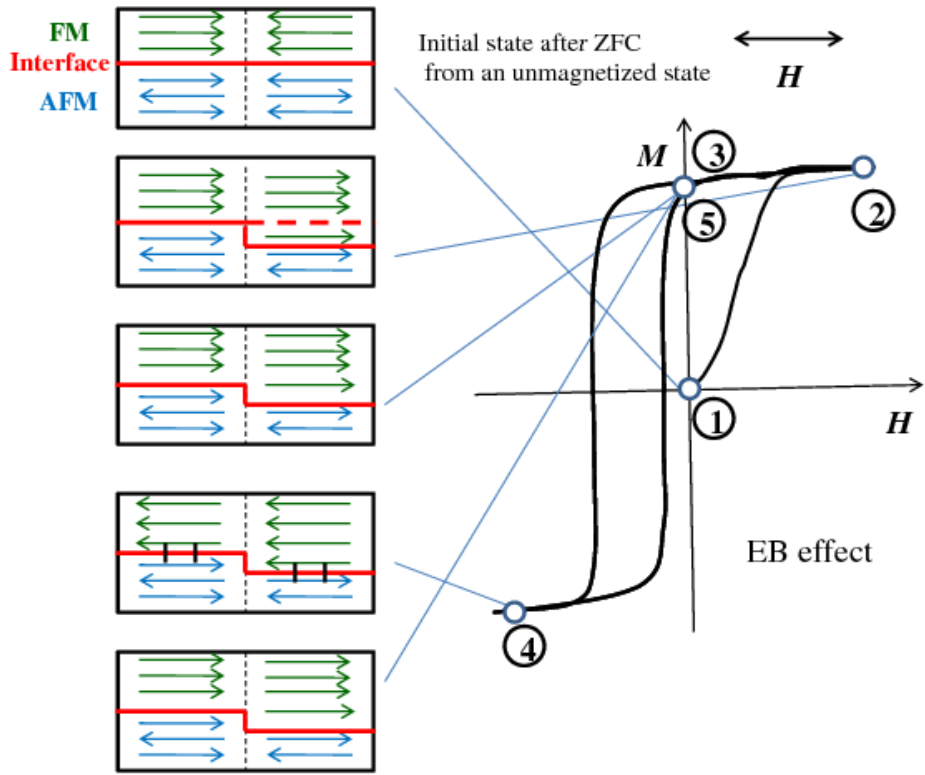

FIG. S2. Hysteresis loop **in the case with a variable interface after** ZFC from an unmagnetized state (right) and schematic diagram of the spin configuration at different stages (lift).

# S3. Compositions of NiMnIn$x$ ($x$ = 11, 12, 13, 14, and 15) determined by EDX analysis

The actual compositions of the nominal compositions Ni$_{50}$Mn$_{50-x}$In$_x$ (NiMnIn$x$) alloys ($x$ = 11, 12, 13, 14, and 15) were determined by energy dispersive x-ray (EDX) analysis [Fig. S3] and are summarized in Table I. The final compositions were obtained by averaging 5 different areas with similar composition in each sample. The actual Ni content in all samples is lower than nominal one due to the loss in the melting process. We used the nominal compositions to represent the samples in the whole text.

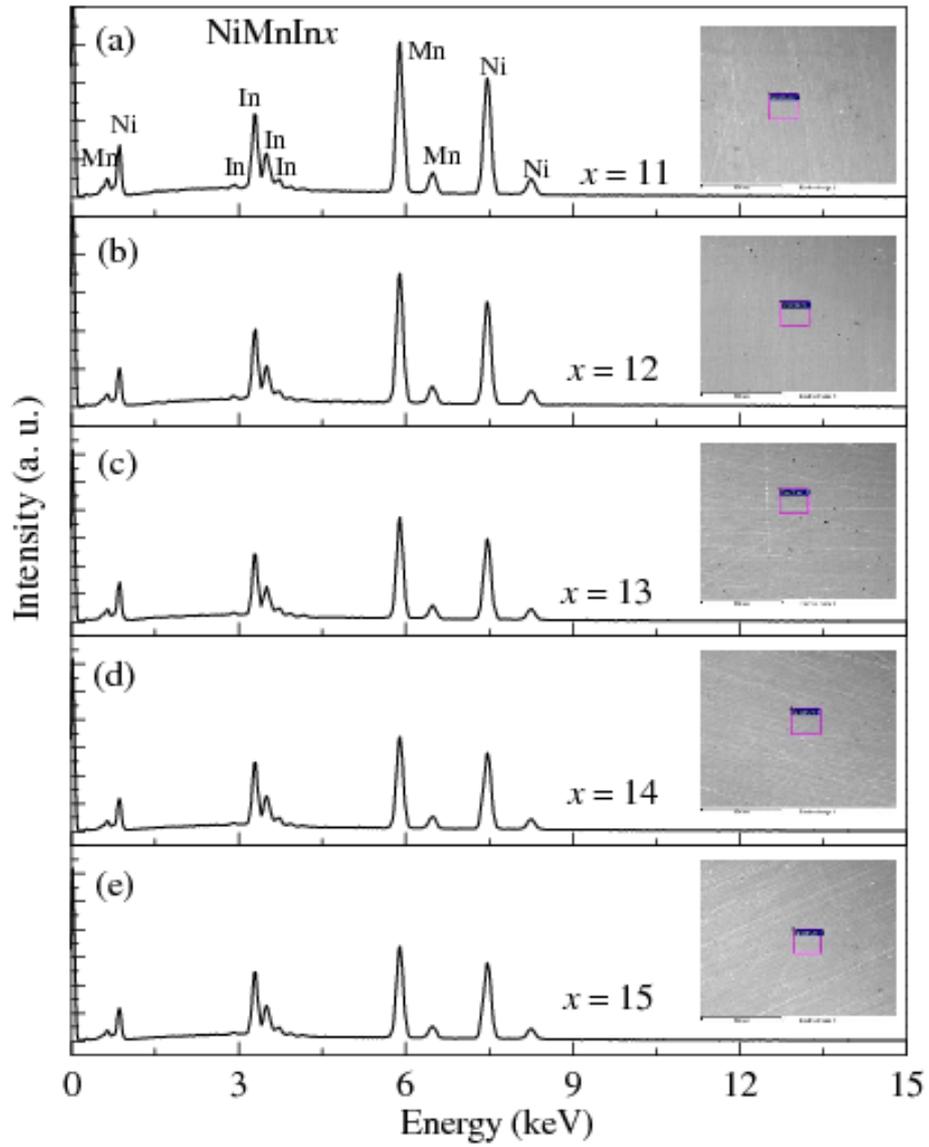

FIG. S3 The EDX spectrums of NiMnIn$x$ ($x$ = 11, 12, 13, 14, and 15). The insets show the scanning electron microscopic pictures. The little rectangles mark the positions where we took the EDX spectrums.

Table I. Compositions of NiMnIn$x$ ($x$ = 11, 12, 13, 14, and 15) determined by EDX analysis.

| $x$ (nominal) | Ni | Mn | In |
|---|---|---|---|
| 11 | 49.2 | 39.4 | 11.4 |
| 12 | 49.6 | 38.4 | 12.0 |
| 13 | 49.0 | 37.5 | 13.5 |
| 14 | 49.2 | 36.5 | 14.3 |
| 15 | 49.0 | 35.5 | 15.5 |

## S4. Crystal structures of NiMnIn$x$ at room temperature

The crystal structures of NiMnIn$x$ at room temperature were determined by *X*-ray diffraction (XRD) on finely powdered specimens [Fig. S4]. Before performing XRD, all samples were annealed at 600°C under high vacuum (~2 × 10$^{-6}$ Torr) for 5 h to remove the residual stress induced during grinding. All the XRD peaks can be indexed for an monoclinic crystal structure, which is consistent with the result observed by Krenke *et al*. [5] for a same nominal composition sample ($x$ = 15). The lattice parameters decrease with increasing the Mn content due to the smaller atomic radius of Mn [Table II].

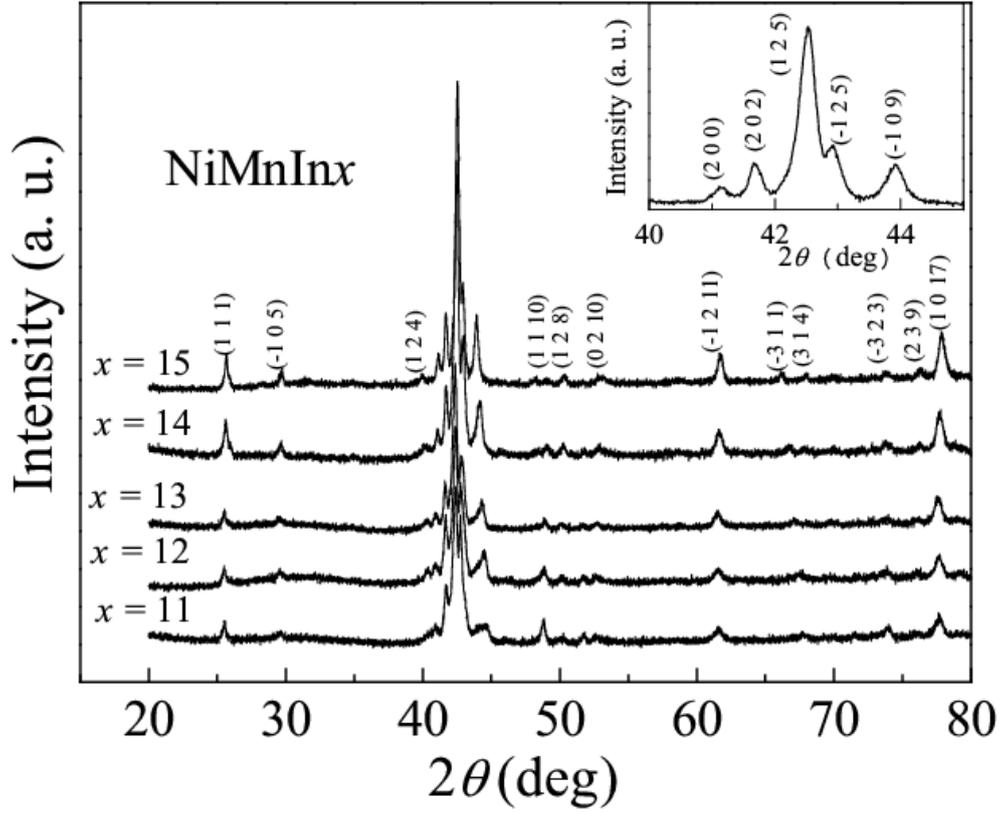

FIG. S4. Powder *X*-ray diffraction patterns at room temperature for NiMnIn*x* alloys(*x* = 11, 12, 13, 14, and 15). The inset shows details in the range $40° \leq 2\theta \leq 44.5°$.

Table II. Lattice parameters of NiMnIn*x* (*x* = 11, 12, 13, 14, and 15) at room temperature.

| *x* | *a* (Å) | *b* (Å) | *c* (Å) | $\beta$ (°) |
|---|---|---|---|---|
| 11 | 4.399 | 5.887 | 21.391 | 87.79 |
| 12 | 4.397 | 5.885 | 21.375 | 87.87 |
| 13 | 4.395 | 5.884 | 21.367 | 87.90 |
| 14 | 4.390 | 5.862 | 21.342 | 88.04 |
| 15 | 4.389 | 5.844 | 21.312 | 88.32 |

## S5. *M*(T) and RM(T) and *M*(H) at 300 K in NiMnIn*x*

*M*(T) curves of NiMnIn*x* were measured under $H$ =100 Oe after ZFC and FC. The remanent magnetization (RM) as a function of temperature in NiMnIn*x* was measured at zero field on heating after the samples were field-cooled under 10 kOe from 300 to 10 K. The RMs decrease with increasing temperature and become zero before reaching 300 K in NiMnIn*x*. The *M*(H) curves at 300 K in NiMnIn*x* are straight lines without any SPM/FM feature, which indicates that their $T_c$s are at lower temperatures. These results make sure that unmagnetized initial states at lower temperatures in these samples can be obtained after ZFC from 300 K.

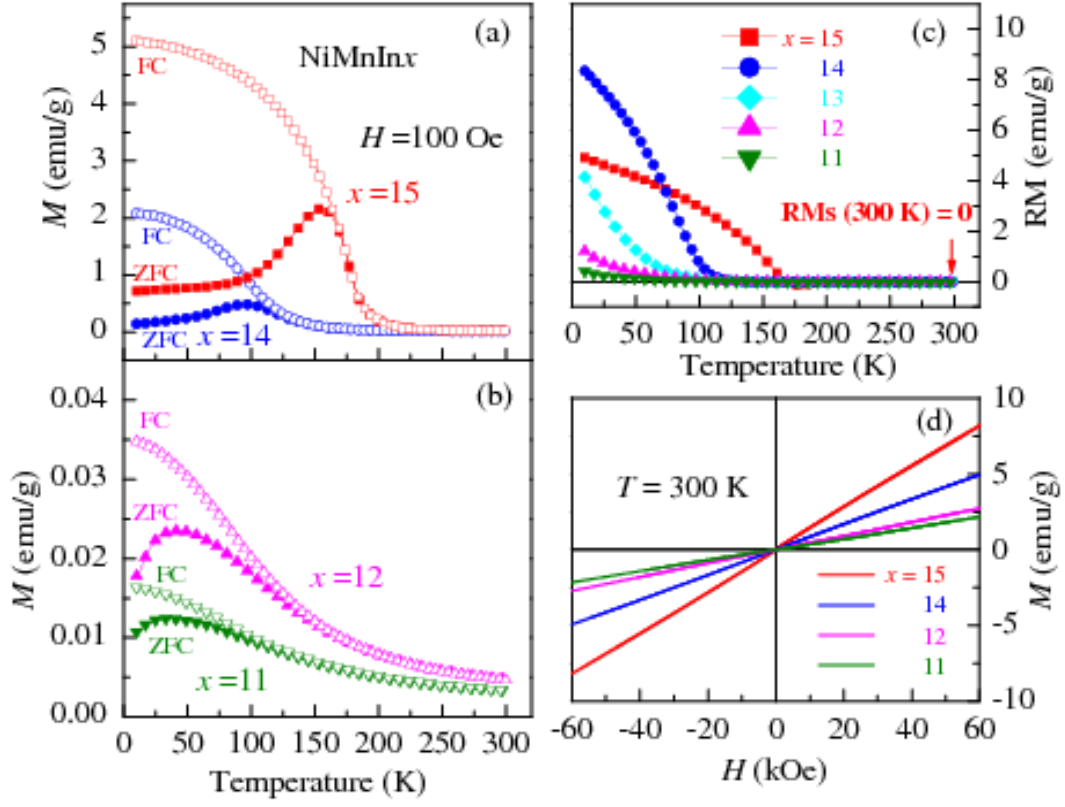

FIG. S5. $M(T)$ measured in $H = 100$ Oe under ZFC and FC in NiMnIn$x$ (a) $x$ = 14, 15. (b) $x$ = 11, 12. (b) RM as a function of temperature for NiMnIn$x$. (c) $M(H)$ curves for NiMnIn$x$ at 300 K.

## S6. Néel temperatures ($T_N$s) of NiMnIn$x$

In NiMn$X$ ($X$ = In, Sn, Sb, Ga *et al.*) alloys, the magnetic transition always accompany with the first-order martensitic transformation. In NiMn$X$ ($X$ = In, Sn, Sb, Ga *et al.*) alloys, the magnetic transition always accompany with the first-order martensitic transformation. This kind of PM-AFM transition has been confirmed by recent neutron polarization analysis in $Ni_{50}Mn_{40}Sn_{10}$ with higher Mn content [6]. Furthermore, FM-AFM transition in $Ni_{50}Mn_{37}Sn_{13}$ with lower Mn content [6] and PM-FM transition in $Ni_{2.19}Mn_{0.81}Ga$ [7] accompanying with the first-order martensitic transformation have also been observed in these systems. Thus, we can use differential scanning calorimetry (DSC), a technique which is sensitive to the first-order phase transition, to determine the Néel temperature of NiMnIn$x$ with higher Mn content in the present study [Fig. S6(a)]. The $T_N$ increases continuously with decreasing In content [Fig. S6(b)], which is consistent with the previous result in NiMn alloy ($T_N \sim 1140$ K) [8].

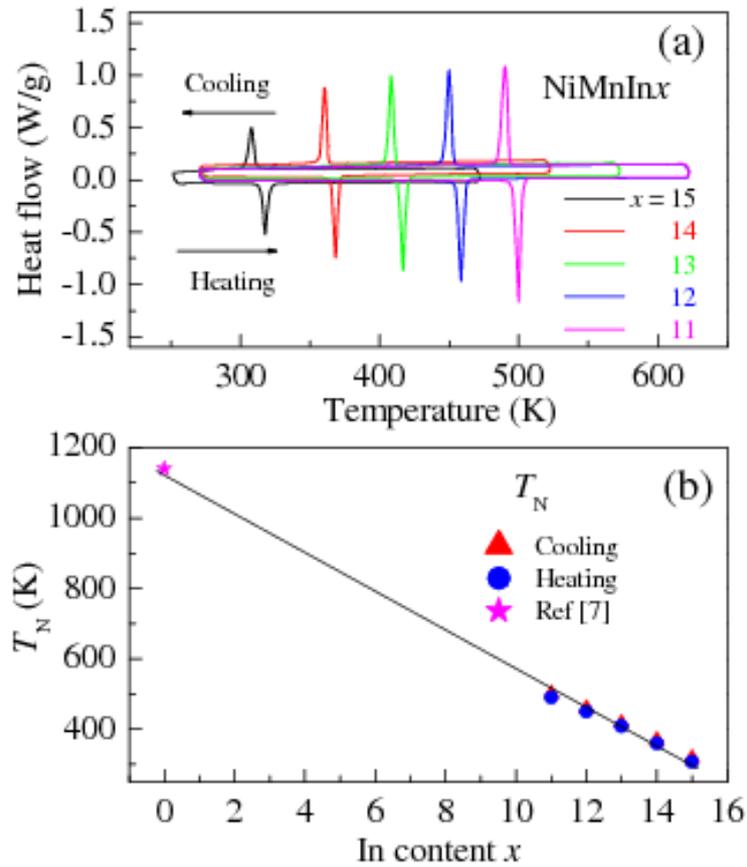

FIG. S6. (a) DSC curves of NiMnIn$x$. The arrows show the heating and cooling directions. (b) $T_N$ as a function of In content.

## S7. CEB effect after FC in NiMnIn$x$

The CEB effect after FC has also been observed in NiMnIn$x$ ($x$ = 11, 12, 13, 14, and 15) (Fig. S7), which is due to the coexistence of AFM and FM phases and interaction at the interface below $T_B$ [9-11].

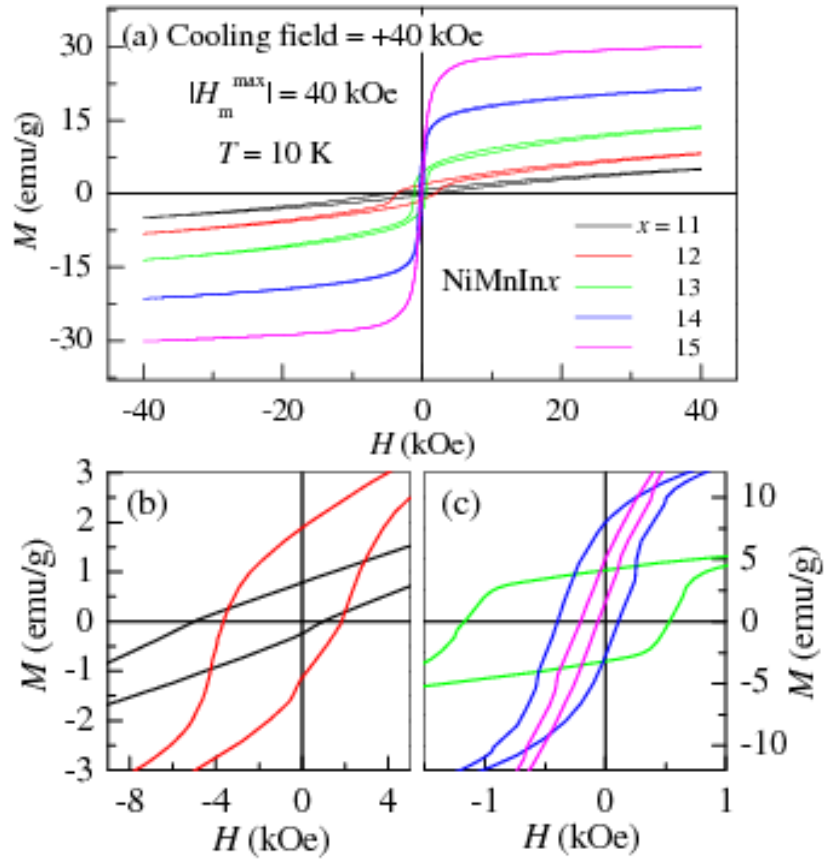

FIG. S7. (a) Hysteresis loops of NiMnIn$x$ at 10 K after FC ($H$ = 40 kOe) from 300 K. The larger scale at low field curves are shown in (b) $x$ = 11, 12, (c) $x$ = 13, 14, 15. The $M$(H) curves shifted along the field axis show the EB effect.

## S8. Memory effect of SSG state in NiMnIn13

To further confirm SSG state at low temperature in NiMnIn13 after ZFC, we used the well defined stop-and-wait protocol to measure its memory effect in the ZFC $M$(T) as shown in Fig. S8, which is an unequivocal signature of SSG behavior [12]. The sample was first ZFC from 300 K to 10 K and then the $M$(T) (reference line, curve 1) was measured during heating under $H$ = 200 Oe. In the stop-and-wait protocol, the sample was ZFC from 300 K to an intermittent stop temperature $T_w$ = 32 K and waited for $10^4$ s followed by further cooling to 10 K. The $M$(T) (curve 2) was then measured under the same conditions as chosen for the reference line. The inset of Fig. S8 shows the difference between curve 2 and 1. The dip at $T_w$ = 32 K clearly shows the memory effect.

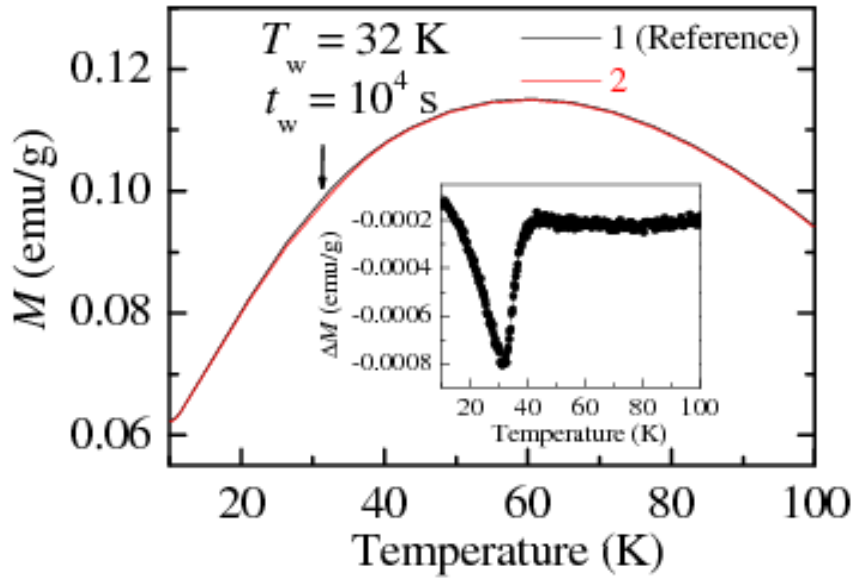

FIG. S8. Temperature dependence of the reference magnetization (curve 1) and of the magnetization with a stop-and-wait protocol at $T_w$ =32 K with waiting time $t_w$ = $10^4$ s

(curve 2), measured in $H$ = 200 Oe. The inset shows the temperature dependence of magnetization difference between curve 2 and 1.

## S9. Hysteresis loops after ZFC from an unmagnetized state and FC in NiMnIn13

An unmagnetized state at 10 K can be obtained after ZFC (the remanent field of magnet is zero) from an unmagnetized state (ZFC starts from an unmagnetized state of the samples). Before each ZFC $M$(H) loop measurement in the present case, there are two methods/conditions have been used: (1) Before each ZFC, the superconductor magnet was demagnetized by oscillating fields at 300 K (in PPMS) or heated up to its superconductor transition temperature (in SQUID, magnet reset) to remove the pinned magnetic flux. (2) The samples were zero-field cooled from 300 K. Since the $T_c$s of all samples are lower than 300 K, the ZFC starts from an unmagnetized state. The magnetization at the starting point of the initial magnetization curve ($H$ = 0) is zero, indicating the initial state at 10 K is an unmagnetized state. That is, the shifted $M$(H) curves (EB effect) are obtained isothermally from an unmagnetized initial state. Furthermore, the shift direction of $M$(H) loop is strongly dependent on the direction of the initial magnetizing field, further indicating the isothermal formation of magnetic unidirectional anisotropy (the origin of EB effect) during the initial magnetization process.

**a. P type $M(H)$ loops measured under different $|H_m^{max}|$ at 10 K after ZFC**

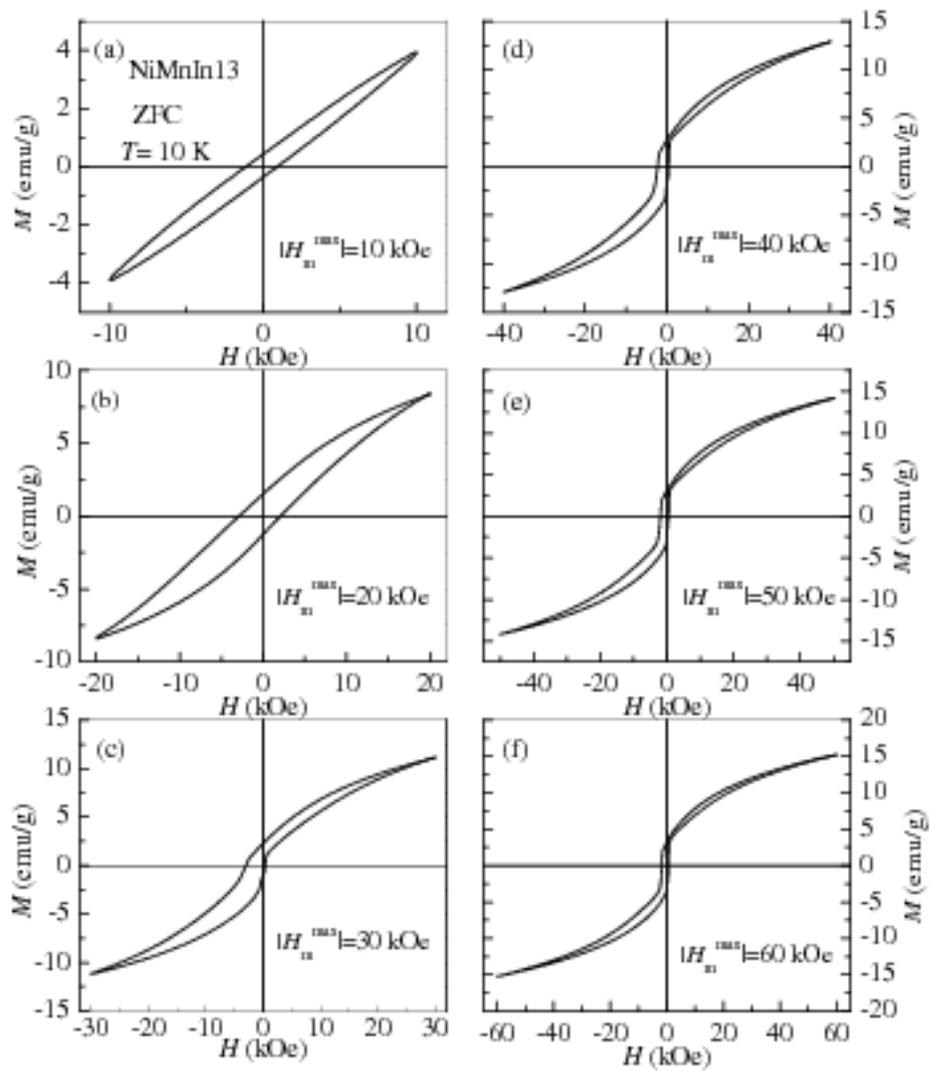

FIG. S9. P type $M(H)$ loops $[0 \to (+H) \to 0 \to (-H) \to 0 \to (+H)]$ measured under different $|H_m^{max}|$s at 10 K after ZFC from 300 K. The initial magnetization curves $[0 \to (+H)]$ are shown in the next figure.

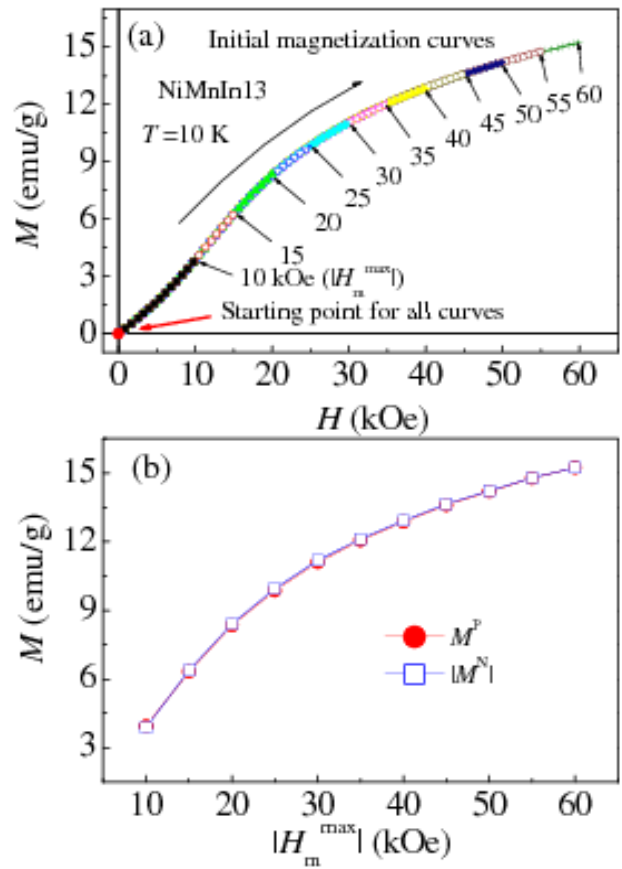

FIG. S10. (a) Initial magnetization curves $[0 \to (+H)]$ for different $|H_m^{max}|$s at 10 K after ZFC from 300 K. (b) The magnetization values in the highest positive ($M^P$) and negative ($|M^N|$) magnetic fields for different $|H_m^{max}|$s.

## b. *M*(H) curves at 10 K after FC

In order to compare results from FC with that from ZFC, we removed the cooling field at 10 K after FC. Then we measured the *M*(H) loop under P and N types, respectively. As expected from the CEB effect after FC, the *M*(H) curve is only dependent on the direction of cooling field [Fig. S11(d)], and it is independent on the direction of the initial magnetization field [Fig. S11(c)]. Furthermore, both the $H_{EB}$ and $H_c$ after ZFC can be larger than those after FC [Fig. S12 (b) and (c)], which indicates EB after ZFC in the present case is not a spontaneous EB [13].

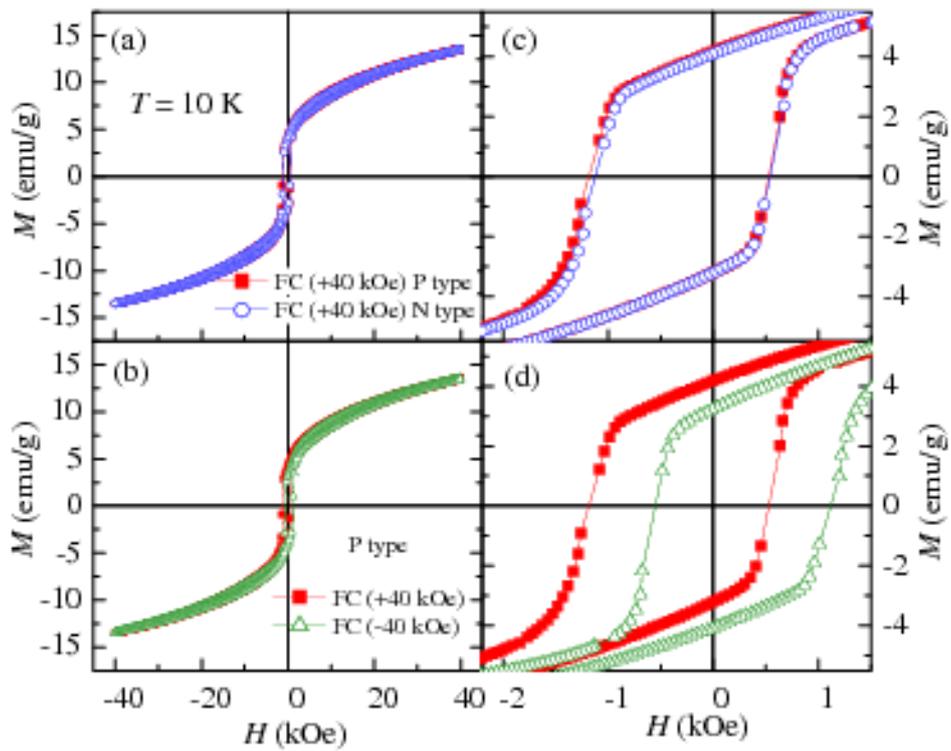

FIG. S11. *M*(H) loops measured at 10 K with $|H_m^{max}|$ = 40 kOe after FC (a) P and N types under the same cooling field (both direction and magnitude), (b) P type under the opposite directions of the same magnitude of the cooling field. (c) and (d) are the larger scale at low field for (a) and (b), respectively.

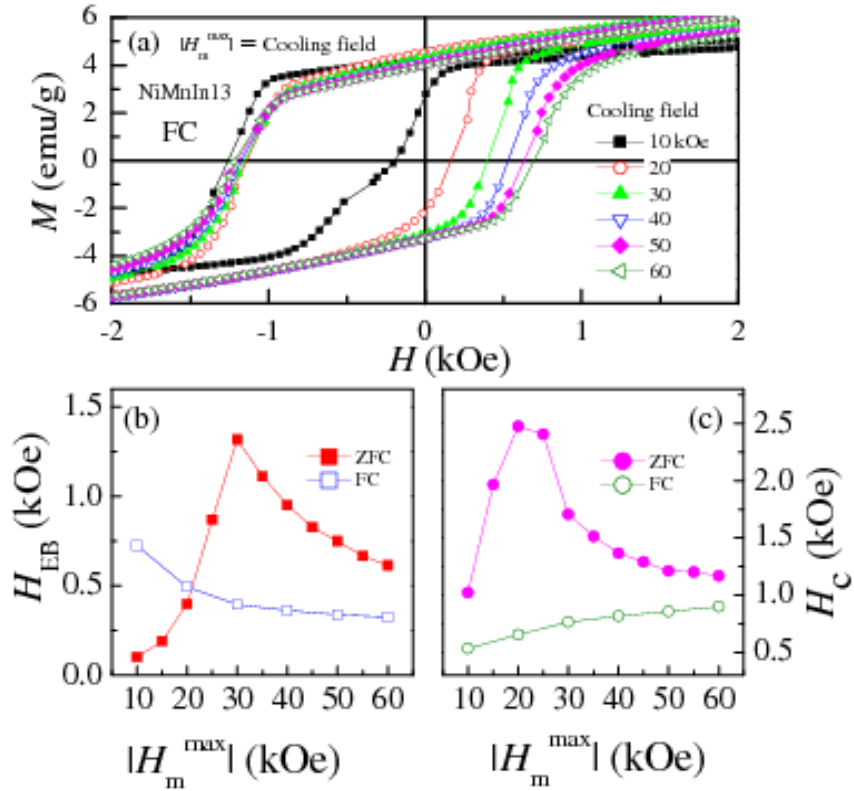

FIG. S12. (a) $M(H)$ loops measured at 10 K after FC from 300 K under different cooling field. The $|H_m^{max}|$ is equal to the cooling field for each curve. For example, cooling field is +60 kOe, sample is cooled with +60 kOe from 300 K to 10 K, then a closed $M(H)$ curve is measured following +60 kOe → 0 → -60 kOe → 0 → +60 kOe. The $H_{EB}$ (b) and $H_c$ (c) as a function of $|H_m^{max}|$ obtained after ZFC an FC.

## c. Strong initial magnetization field direction dependence of EB effect after ZFC

The P and N type $M(H)$ loops measured after ZFC from an unmagnetized state do not overlap and shows a centrally symmetric behavior [Fig. S13], which is unexpected from the CEB systems. This result also cannot be expected from the effect of remanent field of superconductor magnets/remanent magnetization of the samples, in which the direction of hysteresis loop shift is independent on the direction of the initial magnetization field.

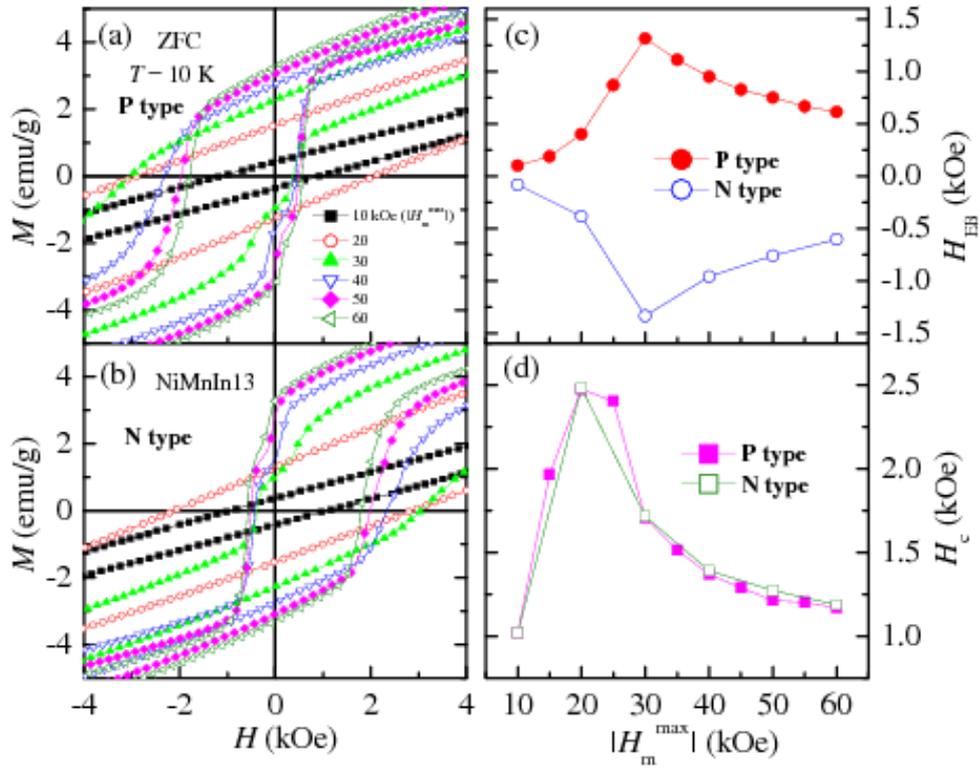

FIG. S13. (a) The low field part of P type $M(H)$ loops $[0 \rightarrow (+H) \rightarrow 0 \rightarrow (-H) \rightarrow 0 \rightarrow (+H)]$ for different $|H_m^{max}|$s at 10 K after ZFC from 300 K shown in Fig. S9. (b) The low field part of N type $M(H)$ loops $[0 \rightarrow (-H) \rightarrow 0 \rightarrow (+H) \rightarrow 0 \rightarrow (-H)]$ for different

$|H_m^{max}|$s at 10 K after ZFC from 300 K. The initial magnetization curves $[0 \to (+H/-H)]$ are not shown. The $H_{EB}$ (c) and $H_c$ (d) as a function of $|H_m^{max}|$ for P an N types $M(H)$ loops.

### d. P and N type *M*(H) loops after ZFC measured by using SQUID

To further rule out the effect from remanent field of superconductor magnet, we measured the $M(H)$ loops after ZFC by using SQUID with magnet reset function. Before each ZFC, the superconductor magnet was heated up to its superconductor transition temperature to remove the pinned magnetic flux (magnet reset). Figure S14 shows the P and N types $M(H)$ loops after ZFC, which are the same as the results obtained by PPMS (Fig. 2 in the main text, before each ZFC, the superconductor magnet was demagnetized by oscillating fields at 300 K).

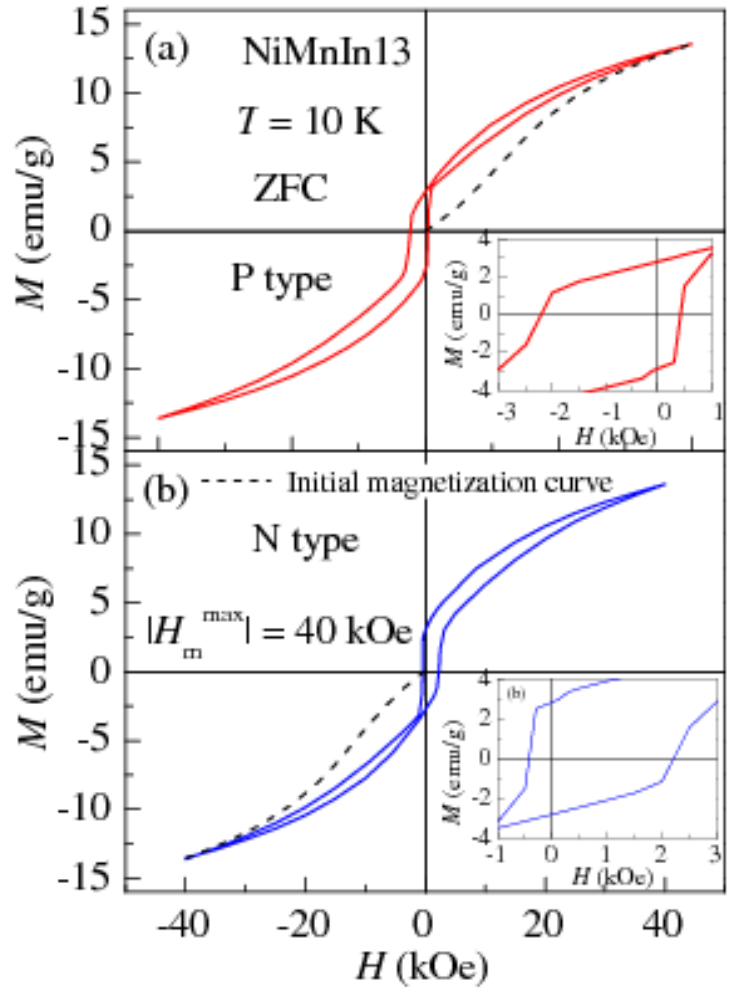

FIG. S14. (a) P and (b) N types $M(H)$ loops at 10 K after ZFC measured by using SQUID. Before each ZFC, the superconductor magnet was heated up to its superconductor transition temperature to remove the pinned magnetic flux (magnet reset). The dash lines show the initial magnetization curves. The insets give the larger scale at the low field.

**e. Temperature dependence of EB after ZFC**

To further confirm this phenomenon after ZFC, we measured the temperature dependence of $H_{EB}$ and $H_c$ for $|H_m^{max}| = 40$ kOe in NiMnIn13 as shown in Fig. S15. The $H_{EB}$ approximately linearly decreases with increasing temperature at low temperatures and gradually disappears around the blocking temperature ($T_B$), at which the $H_c$ reaches its maximum value. These relationships are similar to those in the CEB systems obtained after FC [4]. The AFM anisotropy decreases with the increasing of temperature. Thus, the FM rotation can drag more AFM spins, giving rise to the increase in $H_c$; whereas the AFM can no longer hinder the FM rotation above $T_B$. As a result, the $H_c$ reaches its maximum value and $H_{EB}$ reduces to zero at higher temperatures.

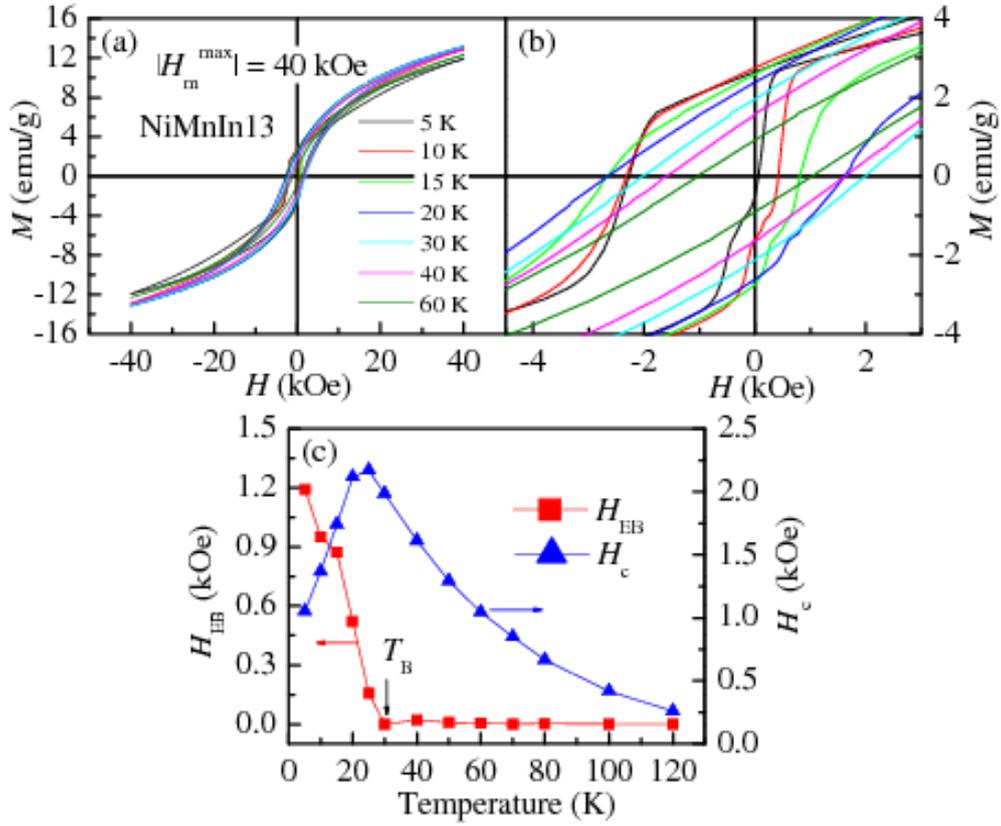

FIG. S15. (a) Temperature dependence of P type $M(H)$ loops with $|H_m^{max}| = 40$ kOe after ZFC from 300 K. (b) The low field part of $M(H)$ loops shown in Fig. S15(a). (c) Temperature dependence of $H_{EB}$ and $H_c$ after ZFC for $|H_m^{max}| = 40$ kOe.

**f. Training effect of EB at 10 K after ZFC**

Figure S16 (a)-(e) show P type $M$ (H) loops measured for 10 consecutive cycles at 10 K after ZFC from 300 K with several selected $|H_m^{max}|$s, and (f) the $H_{EB}$ as a function of number of cycles at 10 K after ZFC for $|H_m^{max}|$ = 15, 25, 35, 45, 55 kOe. There is a large decrease in $H_{EB}$ occurs between $n$ = 1 and $n$ = 2, followed by a more gradual decrease for subsequent loops. The training effect in NiMnIn13 after ZFC is also similar to that in the conventional EB obtained after FC, which composes two distinct mechanisms [14]. One is due to the biaxial anisotropy of the AFM resulting in abrupt single cycle training, while the other is related to the depinning of uncompensated AFM spins resulting in a gradual decrease for subsequent loops.

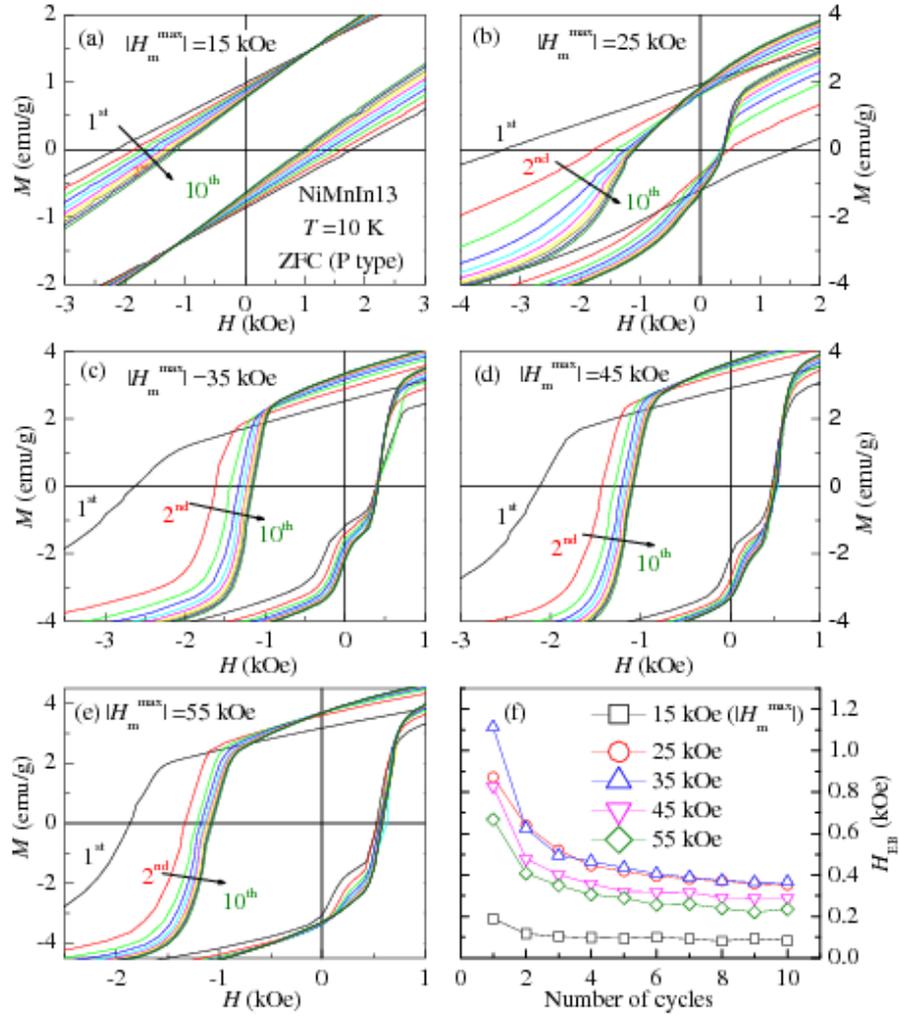

FIG. S16 (a)-(e) P type $M$ (H) loops measured for 10 consecutive cycles at 10 K after ZFC from 300 K with several selected $|H_m^{max}|$s. (f) $H_{EB}$ as a function of number of cycles at 10 K after ZFC for $|H_m^{max}|$ = 15, 25, 35, 45, and 55 kOe.

## S10. Further supports to the model shown in Fig. 3 in the main text

**a. Anomalous remanent magnetization dependence of EB effect**

Miltényi *et al.* [15] reported that the EB effect can be well tuned by cooling in zero field from different remanent magnetization (RM) states in the CEB systems. The $H_{EB}$ increases with increasing RM and the sign of $H_{EB}$ changes when changing the sign of RM. This allows one to select the desired value of $H_{EB}$ through a simple cooling procedure after device preparation, which shows a very technological importance. Here, we show an anomalous RM dependence of EB effect in NiMnIn13. Both the value of exchange bias field and its sign can be tuned by the amplitude of RM without changing its sign. This tunability is strongly dependent on the direction of the initial magnetization field for the hysteresis loop measurements. These results can be explained well by isothermal field-induced transition from SSG to SFM state as shown in Fig. 3 in the main text.

To obtain the different magnetization states in NiMnIn13, we applied different positive fields on sample at 35 K (above $T_B$ = 30 K) and removed them as shown in Fig. S17(a). The result shows that this procedure can establishes different RMs by applying different fields. Then, the sample was zero-field cooled to 10 K. After that, the P and N types *M*(H) loops were measured, respectively. During the ZFC from 35 K to 10 K, the RM reduces a little. Figure S17(b) shows the RMs at 10 K as a function of applied field at 35 K. The RM at 10 K can be changed continuously by applying different fields at 35 K. The P and N type *M*(H) loops were measured with $|H_m^{max}|$ = 40 kOe for each RM state, respectively. The equal magnetization values in the highest positive and negative magnetic fields indicate the shifted loop is not a nonsymmetrical minor hysteresis loop. It is worth noting that the P and N types $M$(H) loops shift to the opposite field direction at small RM state [+0.37 emu/g, Fig. S18(c)], while shift to the same field direction at large RM state [+1.59 emu/g, Fig. S18(d)]. There is a difference between the first cycle of P and N types $M$(H) loops for larger RM state [Fig. S18(d)]. The second cycle of P type $M$(H) loop is the same as the first cycle of N type $M$(H) loop. These results may be due to the large training effect in NiMnIn13. The first cycle of N type $M$(H) loop, the direction of the initial magnetization field direction is opposite to the RM direction, is actually obtained after a magnetization reversal. Namely, it should be regarded as a second cycle hysteresis loop.

Figure S19 shows the RM dependence of $H_{EB}$ at 10 K obtained from the first and second cycles of P and N types $M$(H) loops. The large difference between two cycles of P or N types $M$(H) loops is due to the large training effect in the system. For N type, not only the value of $H_{EB}$ but also its sign can be tuned by changing the value of RM. For P type, only the value of $H_{EB}$ decreases with increasing RM at the first cycle and it is almost a constant at the second cycle. These results are different from the RM dependence of EB effect in the previous CEB systems: the $H_{EB}$ decreases monotonically with decreasing RM and changes its sign when the sign of RM is changed [15]. Note that there is a large EB effect at zero RM state, which is attributed to the field-induced transition from SSG to SFM states in the main text.

Here, we use the model shown in Fig. 3 in the main text to explain the anomalous RM dependence of EB effect in NiMnIn13 (Fig. S20). Note that the diagram of Fig. S20 only shows the change of SFM volume fraction and magnetization direction at different stages, which is different from the diagram of Fig. 3 in the main text showing the process of isothermal field-induced SSG to SFM transition. The initial magnetic state is SSG state at 35 K after ZFC, which is similar to that at 10K. When a large field $+H_2 > +H_{35K}^{crit}$ ($H_{35K}^{crit}$ is a critical field at which SSG state just completely transforms to SFM state at 35 K, which is different from the critical field at 10 K) is applied, all of the SSG state will transform to SFM state. After removal of field, there is a large RM of SFM. This RM persists to lower temperature after ZFC. The white arrows in Fig. S20 show the *net* RM parallel to the direction of field at 10K. The non-zero net positive RM is pinned by the AFM matrix below $T_B$ (= 30 K), which will produce negative EB effect and the EB effect is independent on the direction of the initial magnetization field. This is similar to that observed in the previous CEB systems [15].

When a small field $+H_1 < +H_{35K}^{crit}$ is applied at 35 K, only part of the SSG state can transform to the SFM state, resulting in a small positive RM of SFM. The mixed SSG and SFM states persist to 10 K after ZFC from 35 K. Note that the larger magnetic field can still transform the remanent SSG to SFM states at 10 K and the induced SFM magnetization direction is determined by the direction of the initial magnetization field. For P type measurement with the positive initial magnetization field [$+H_3$ (+40 kOe) > $+H^{crit}$ (30 kOe)], the SSG state transforms to SFM state with positive RM. That is, the net RM direction is always along the positive field direction for P type $M(H)$ loops regardless of the initial state (pure SFM or mixed SSG and SFM states). The positive RM will produce negative EB, which is consistent with the experiment results [Fig. S18(c)]. While for N type measurement with the negative initial magnetization field ($|-H_3|$ (-40 kOe) > $+H^{crit}$), the remanent SSG state transforms to SFM state with negative RM. Therefore, the net RM along the field direction is dependent on the values of the positive RM of the initial SFM state and negative RM of newly induced SFM state. The magnetometry measures the average coupling over the whole interface area in the sample. The different values of net positive or negative RM at 10 K can be obtained from N type $M$(H) loops through applying different magnetic fields at 35 K, which will produce negative or positive EB effect with different value of $H_{EB}$. In a word, the magnitude of $H_{EB}$ and its sign can be tuned effectively by the value of RM in the initial state.

In summary, we observed an anomalous RM dependence of EB effect in NiMnIn13. It can be explained well by field-induced transition from SSG to SFM state as shown in Fig. 3 in the main text. The RM at 10 K in the initial state not only reflects the magnetization state of the sample, but also reflects the ratio of SSG to SFM state transition at 35 K.

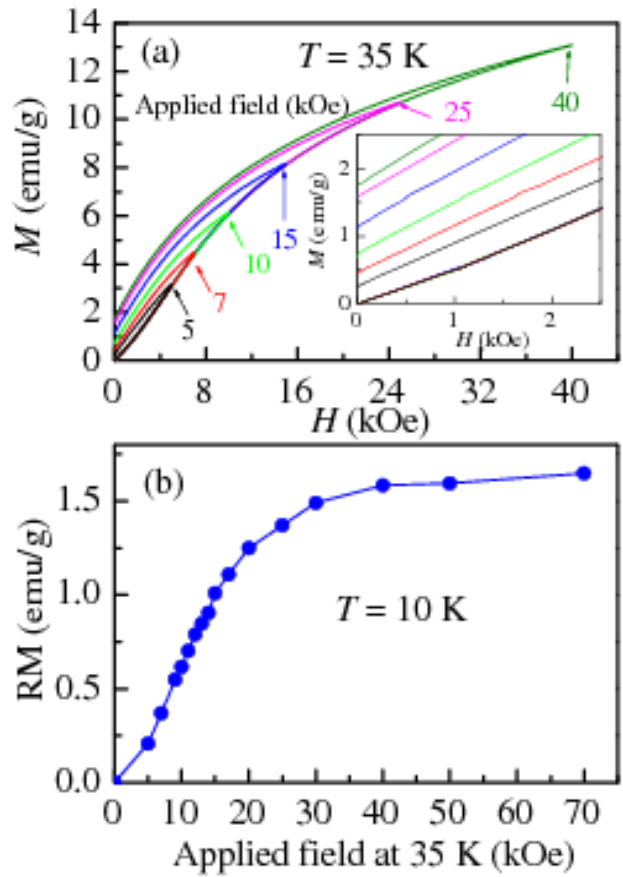

FIG. S17. (a) Isothermal magnetic-field dependence of magnetization for different applied fields at 35 K. The inset: a larger scale at low. (b) The RM at 10 K as a function of applied field at 35 K.

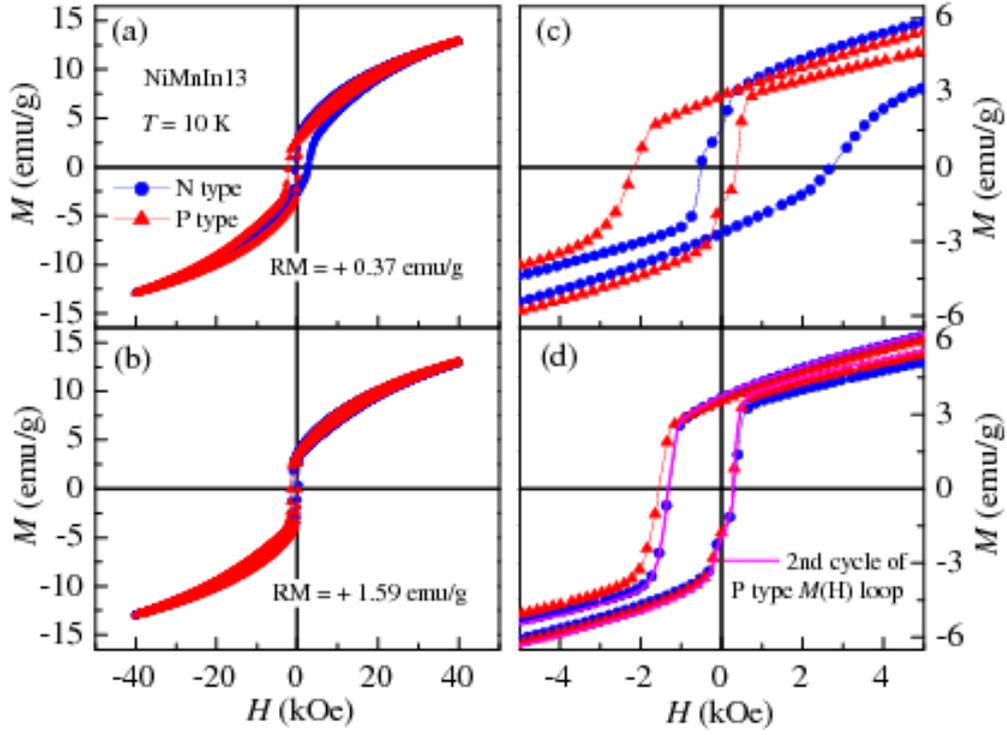

FIG. S18. P and N types $M(H)$ loops ($|H_m^{max}| = 40$ kOe) at 10 K measured after cooling in zero field from 35 K with different RM states, (a) RM = + 0.37 emu/g, (b) RM = + 1.59 emu/g. (c) and (d) are the larger scale at low field in (a) and (b), respectively. The magenta solid line in Fig. 2(b) shows the second cycle of P type $M(H)$ loop.

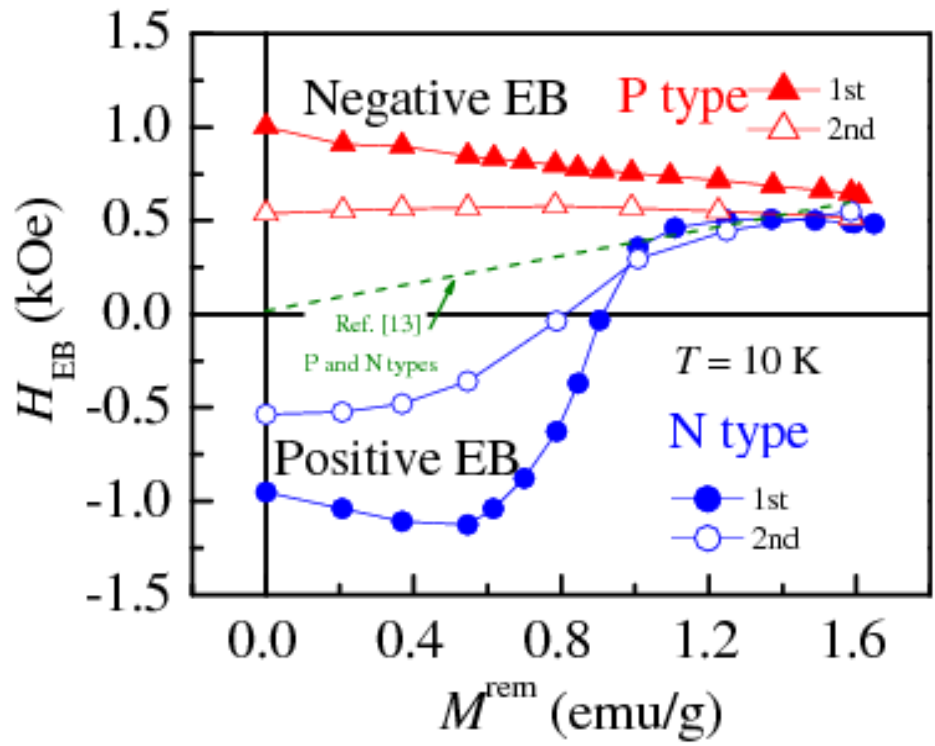

FIG. S19. RM dependence of $H_{EB}$ at 10 K obtained from first and second cycles of P and N types $M(H)$ loops.

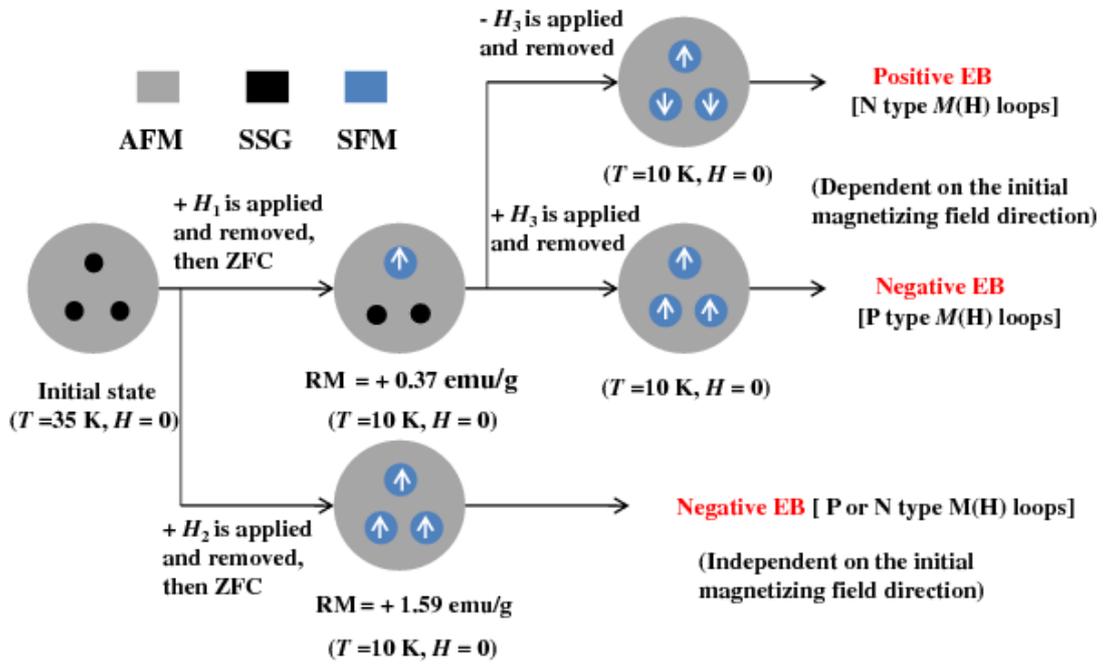

FIG. S20. Schematic diagrams of SSG and SFM states embedded in an AFM matrix at different stages. The white arrows represent the direction of net RM at SFM state parallel to the direction of magnetic field. This diagram only shows the change of SFM volume fraction and magnetization direction at different stages, which is different from the diagram of Fig. 3 in the main text showing the process of isothermal field-induce SSG to SFM transition.

**b. Isothermal tuning of EB after ZFC from an unmagnetized state**

Based on the results shown in Fig. S20, the similar results should be obtained *isothermally* if the initial state is at 10 K after ZFC from an unmagnetized state. In order to check this point, we applied different positive fields ($H^{applied}$) and removed them at 10 K after ZFC from an unmagnetized state, then measured the P and N types $M$(H) loops with $|H_m^{max}|$ = 40 kOe, respectively. As shown in Fig. S21, the P and N types $M$(H) loops shift to the opposite direction at small applied field (+10 kOe), while shift to the same direction at larger applied field (+40 kOe). Figure S22 shows the $H^{applied}$ dependence of $H_{EB}$ at 10 K obtained from P and N types $M$(H) loops. For N type, not only the value of $H_{EB}$ but also its sign can be tuned by changing the magnitude of the applied field. While for P type, only the value of $H_{EB}$ decreases with increasing $H^{applied}$. These results are similar to the anomalous RM dependence of EB effect, which can be explained within the Fig. S20. The isothermal tuning of EB after ZFC from an unmagnetized state in the present case is strongly dependent on the direction of the initial magnetization field, which cannot be expected in the previous CEB system with the similar isothermal tuning effect [16, 17]. According to the field-induced transition from SSG to SFM state as shown in Fig. 3 in the main text, the different positive applied fields produce the different volume fractions of SFM state with positive direction unidirectional anisotropy. After that, for the P type $M$(H) measurement, the positive initial magnetization field [+40 kOe > $H^{crit}$ (30 kOe), SSG state will completely transform to SFM state] will transform remanent SSG state to SFM with positive direction unidirectional anisotropy. So the EB is always negative EB for P type measurement. While for N type $M$(H) measurement, the negative initial magnetization field (- 40 kOe > $H^{crit}$) will transform remanent SSG state to SFM with negative direction unidirectional anisotropy. The direction of net unidirectional anisotropy is dependent on the magnitude of $H^{applied}$ before $M$(H) measurement. That is why the $H_{EB}$ (both value and sign) from N type $M$(H) is strongly dependent on the magnitude of the positive $H^{applied}$.

In order to further confirm isothermal field-induced transition from SSG to SFM state, we measured $M$(H) loops with smaller $|H_m^{max}|$ = 10 kOe after isothermally applying different fields at 10 K. The 10 kOe can only transform little SSG to SFM state at 10 K. So we use 10 kOe measurement field to check the magnetic state changed or not after applying different fields. The sample was first zero-field cooled from 300 to 10 K. Then different positive fields were applied and reduced to +10 kOe. After that, $M$(H) loops were measured following +10 kOe → 0 → -10 kOe → 0 → +10 kOe (1$^{st}$ cycle). Other than the phenomenon of EB effect, the shape of $M$(H) loop was also changed by applying different fields isothermally (Fig. S23). For $H^{applied}$ ≥ 30 kOe, there is an FM-type $M$(H) loop, which is consistent with the result in the main text that $H^{crit}$ = 30 kOe at 10 K. The difference of magnetization between two cycles, only observed at +10 kOe → 0 → -10 kOe branch, may originate from the change of AFM bulk spin structure by larger field and the changed AFM bulk spin structure can recover after field reversal.

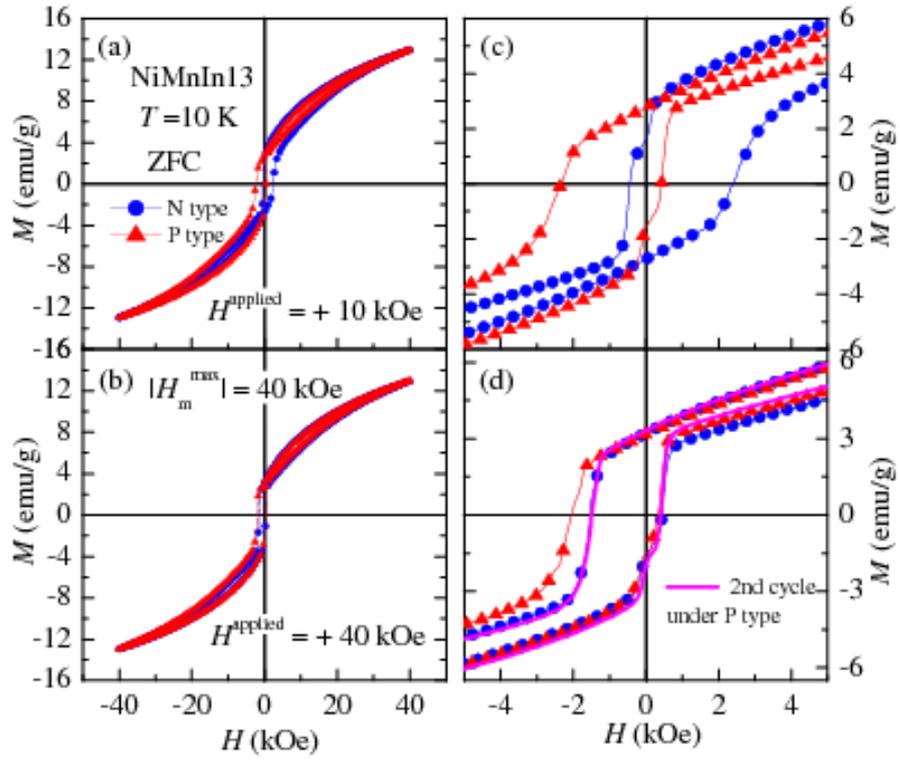

FIG. S21. P and N types $M(H)$ loops ($|H_m^{max}|$ = 40 kOe) at 10 K measured after different positive fields were applied and removed isothermally (a) $H^{applied}$ = + 10 kOe, (b) $H^{applied}$ = + 40 kOe. (c) and (d) are the larger scale at low field in (a) and (b), respectively. The magenta solid line in (d) shows the second cycle of P type $M(H)$ loop.

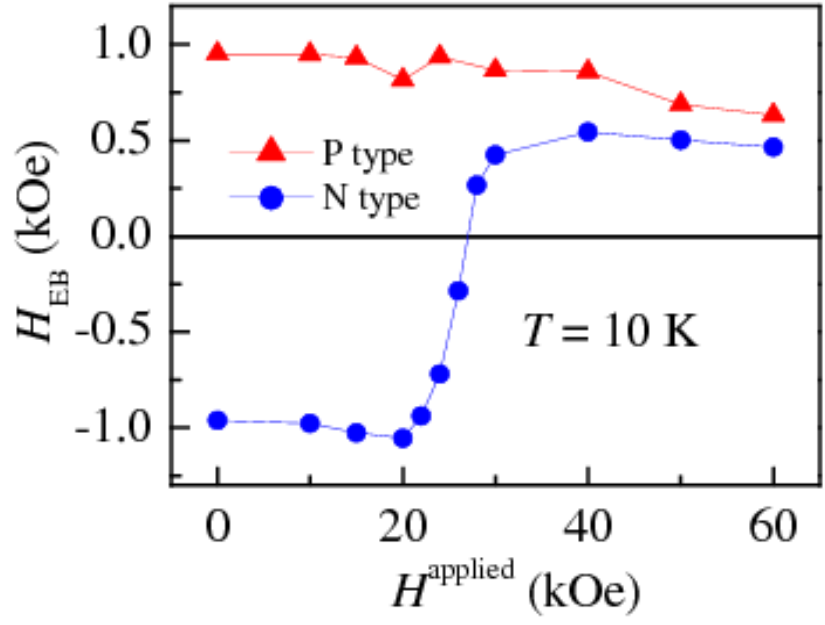

FIG. S22. $H^{\text{applied}}$ dependence of $H_{\text{EB}}$ at 10 K obtained from P and N types $M(H)$ loops.

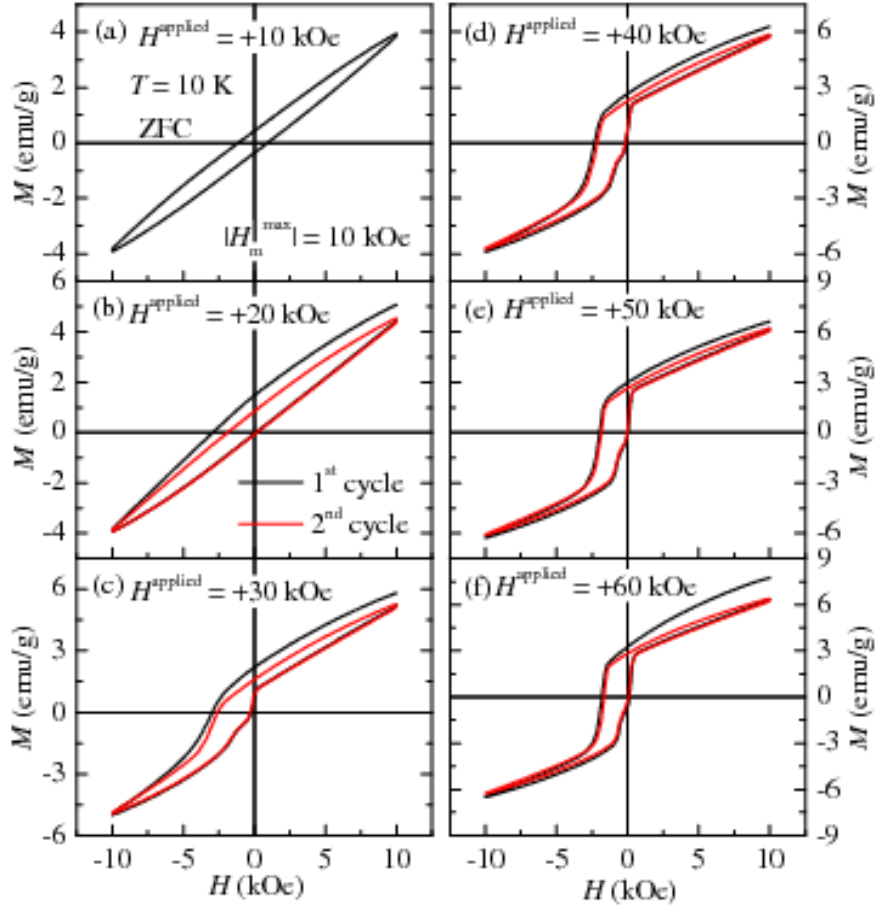

FIG. S23. $M(H)$ loops ($|H_m^{max}| = 10$ kOe) at 10 K measured after applying different positive fields ($H^{applied}$s) isothermally.

**c. Strong cooling field dependence of the CEB effect in NiMnIn13**

In the previous CEB systems, the value of $H_{EB}$ obtained after FC can be tuned by the magnitude of cooling field ($H^{FC}$) [18, 19]. In a lower $H^{FC}$ range, the $H_{EB}$ increases with increasing of $H^{FC}$ due to the saturation of the FM layer magnetization [18]. In a higher $H^{FC}$ range far above $H_c$, if the coupling at the FM/AF interface is FM, $H_{EB}$ is not further affected by $H^{FC}$[18]. However, if the coupling at the interface is AF, $H_{EB}$ is tuned in a large field range and even changes its sign under lager $H^{FC}$, which is ascribed to the competition between the exchange energy and the Zeeman energy [1, 19]. The bulk AFM spin structure has also been shown to play a crucial role on EB effect [20]. Moreover, the value of $H_{EB}$ can also be tuned by the $H^{FC}$ through changing the thickness of FM layer in a spontaneous lamellar FM/AFM phase separated material [21].

We measured the $M(H)$ loops ($|H_m^{max}|$ = 20 kOe) at 10 K after FC from 300 K under different $H^{FC}$s [Fig. S25(a)]. The sample was first cooled with $+H$ from 300 to 10 K, then the magnetic field was reduced/increased to +20 kOe at 10 K isothermally, after that a closed $M(H)$ loop was measured following +20 kOe → 0 → -20 kOe → 0 → +20 kOe (1$^{st}$ cycle). Figure S24(a) shows the 1$^{st}$ and 2$^{nd}$ cycles $M(H)$ loops measured at 10 K after FC ($H^{FC}$ = +60 kOe) from 300 K. $M^{ini}$ is the magnetization under +20 kOe after reducing from +60 kOe. $M^P$ and $M^N$ are the magnetizations under +20 kOe and − 20 kOe in the second cycle, respectively. Figure S24(b) shows the $M^{ini}$, $M^P$ and $|M^N|$ as a function of $H^{FC}$ at 10 K. From these curves we can make the following observation: (1) A difference between $M^{ini}$ and $M^P$ at $H^{FC}$ > 10 kOe, which is related to the change of AFM bulk spin structure by cooling field, (2) The $M^P(=|M^N|)$ increases with the increasing of $H^{FC}$ up to 40 kOe, then decreases slowly at larger $H^{FC}$. The $H_{EB}$ increases steeply with increasing $H^{FC}$ up to a maximum at $H^{FC}$ = 1kOe, then decreases with increasing $H^{FC}$, which is different from that in the CEB systems [Fig. S26(a)] [18]. The $H_{EB}$ can be changed hugely from 1658 Oe for $H^{FC}$ = 1 kOe to 288 Oe for $H^{FC}$ = 80 kOe. The difference of $H_{EB}$ between two cycles shows the training effect of CEB in NiMnIn13.

Considering the large $H^{FC}$ dependence of $M^P$ (= $|M^N|$) at $H^{FC}$ < 40 kOe, the decrease of $H_{EB}$ with the increasing of $H^{FC}$ at this range may partially come from the increase of FM phase volume fraction by cooling field [21]. This result can be obtained from model shown in Fig. 3 in the main text. The SPM/FM domain size increases with increasing applied field. If the sample is cooled down under different fields from higher temperature, the SPM/FM domain size is dependent on the magnitude of the applied field. After further cooling, the increase of interaction among domains transforms SSG to SFM state. That is, the saturation of SFM (volume fraction) is strongly dependent on the cooling field. However, the linear relationship of $H_{EB}$ as a function of $1/M^P$ is only satisfied at 1 kOe < $H^{FC}$ < 10 kOe range [Fig. S26(b)].

The cooling field dependence of $H_{EB}$ in the present case can be divided four zones [Fig. S26(c)]: (1) 0 < $H^{FC}$ < 1 kOe, the $H_{EB}$ increases with increasing of $H^{FC}$ due to the saturation of the FM layer magnetization [18], (2) 1 kOe < $H^{FC}$ < 10 kOe, the $H_{EB}$ decreases with increasing of $H^{FC}$ due to the increasing of FM volume fraction [21], (3) 10 kOe < $H^{FC}$ < 40 kOe, the $H_{EB}$ decreases with increasing of $H^{FC}$ due to the increasing of FM volume fraction and the change of AFM bulk spin structure, (4) 40 kOe < $H^{FC}$ < 80 kOe, the $H_{EB}$ decreases with increasing of $H^{FC}$ due to the change of AFM bulk spin structure [20]. The $H_{EB}$ obtained after ZFC decreases with increasing $|H_m^{max}|$ at large field ($|H_m^{max}|$ > 30 kOe) may come from the same reason.

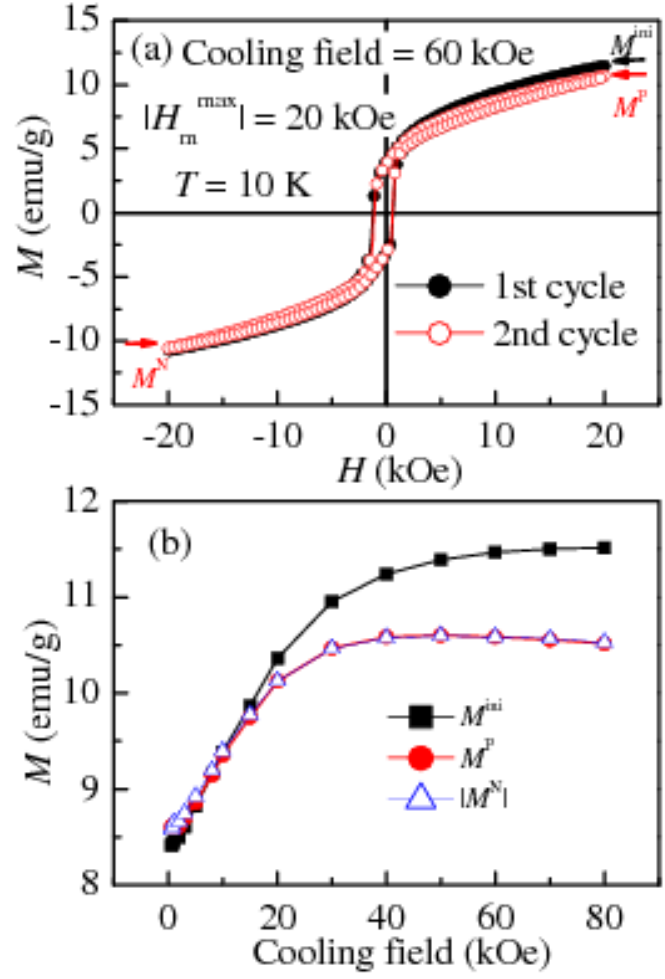

FIG. S24. (a) 1st and 2nd cycles $M(H)$ loops ($|H_m^{max}| = 20$ kOe) measured at 10 K after FC ($H^{FC} = +60$ kOe) from 300 K. The sample is first cooled with +60 kOe from 300 K to 10 K, then the magnetic field is reduced to +20 kOe at 10 K isothermally, after that a closed $M(H)$ curve is measured following +20 kOe → 0 → -20 kOe → 0 → +20 kOe (1st cycle). (b) $M^{ini}$, $M^P$ and $|M^N|$ as a function of $H^{FC}$ at 10 K.

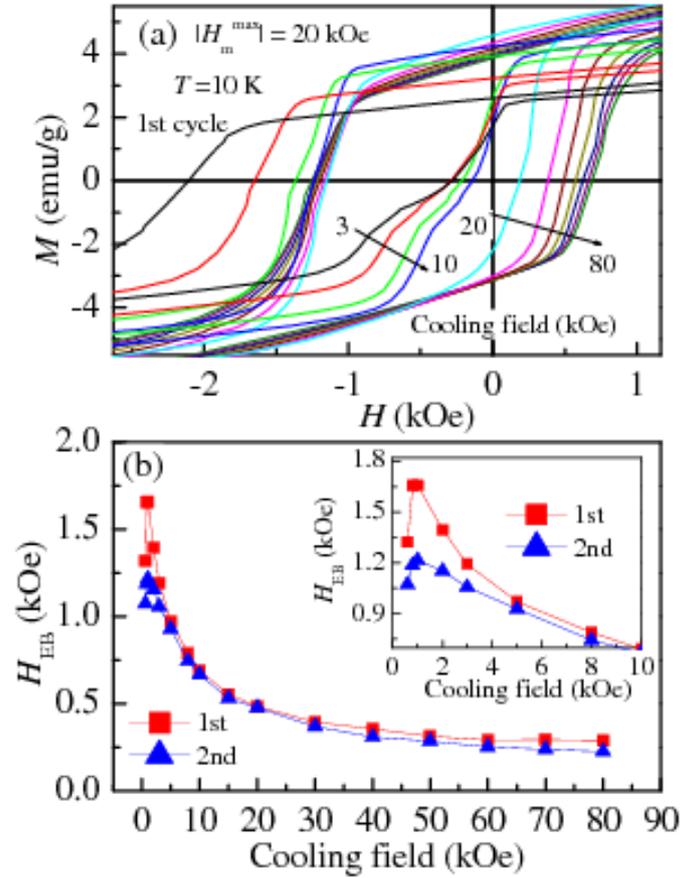

FIG. S25. (a) $M(H)$ loops (1st cycles, $|H_m^{max}| = 20$ kOe) measured at 10 K after FC from 300 K under different $H^{FC}$s. The sample is first cooled with $+H$ from 300 K to 10 K, then the magnetic field is reduced/increased to $+20$ kOe at 10 K isothermally, after that a closed $M(H)$ curve is measured following $+20$ kOe $\rightarrow 0 \rightarrow -20$ kOe $\rightarrow 0 \rightarrow +20$ kOe (1st cycle). (b) $H_{EB}$ as a function of $H^{FC}$ for 1st and 2nd cycles. The inset gives a larger scale at smaller cooling fields.

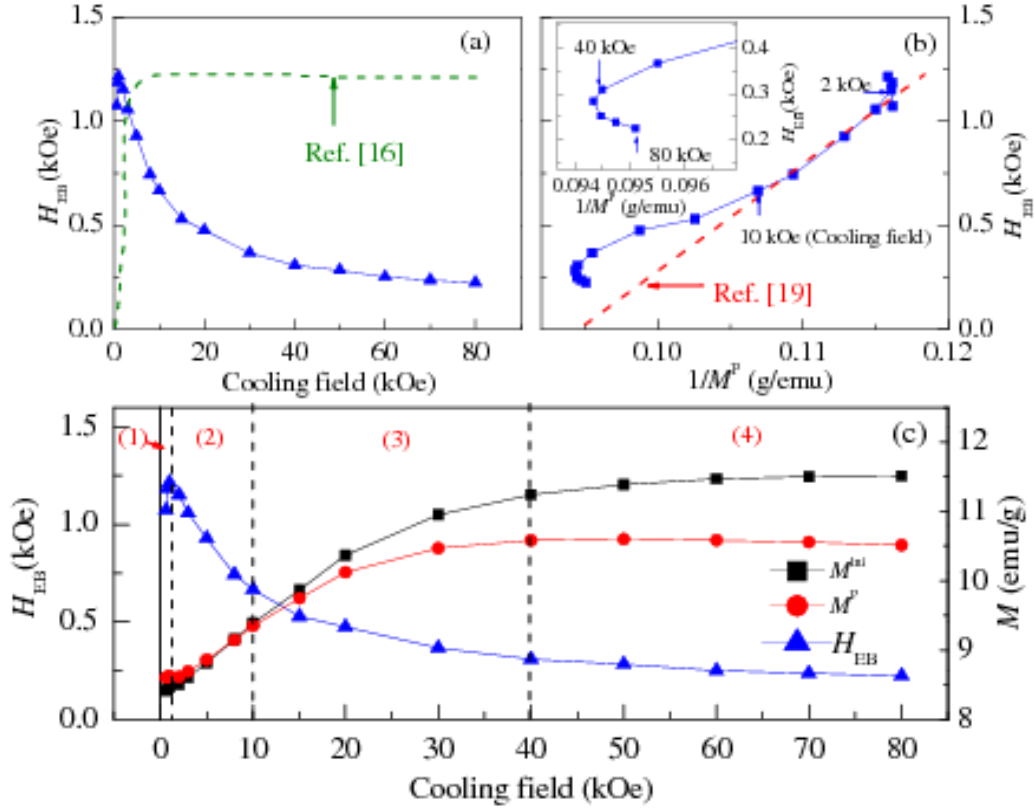

FIG. S26. (a) Compare $H_{EB}$ as a function of $H^{FC}$ with that in Ref. [18]. (b) Compare $H_{EB}$ as a function of $1/M^P$ with that in Ref. [21]. (c) The cooling field dependence of $H_{EB}$ in the present case can be divided four zones: (1) $0 < H^{FC} < 1$ kOe, the $H_{EB}$ increases with increasing of $H^{FC}$ due to the saturation of the FM layer magnetization [18], (2) 1 kOe $< H^{FC} < 10$ kOe, the $H_{EB}$ decreases with increasing of $H^{FC}$ due to the increasing of FM volume fraction [21], (3) 10 kOe $< H^{FC} < 40$ kOe, the $H_{EB}$ decreases with increasing of $H^{FC}$ due to the increasing of FM volume fraction and the change of AFM bulk spin structure, (3) 40 kOe $< H^{FC} < 80$ kOe, the $H_{EB}$ decreases with increasing of $H^{FC}$ due to the change of bulk AFM spin structure [20].

# S11. Hysteresis loops after ZFC from an unmagnetized state in NiMnIn$x$ ($x$ = 11, 12, 14, and 15)

## a. NiMnIn11

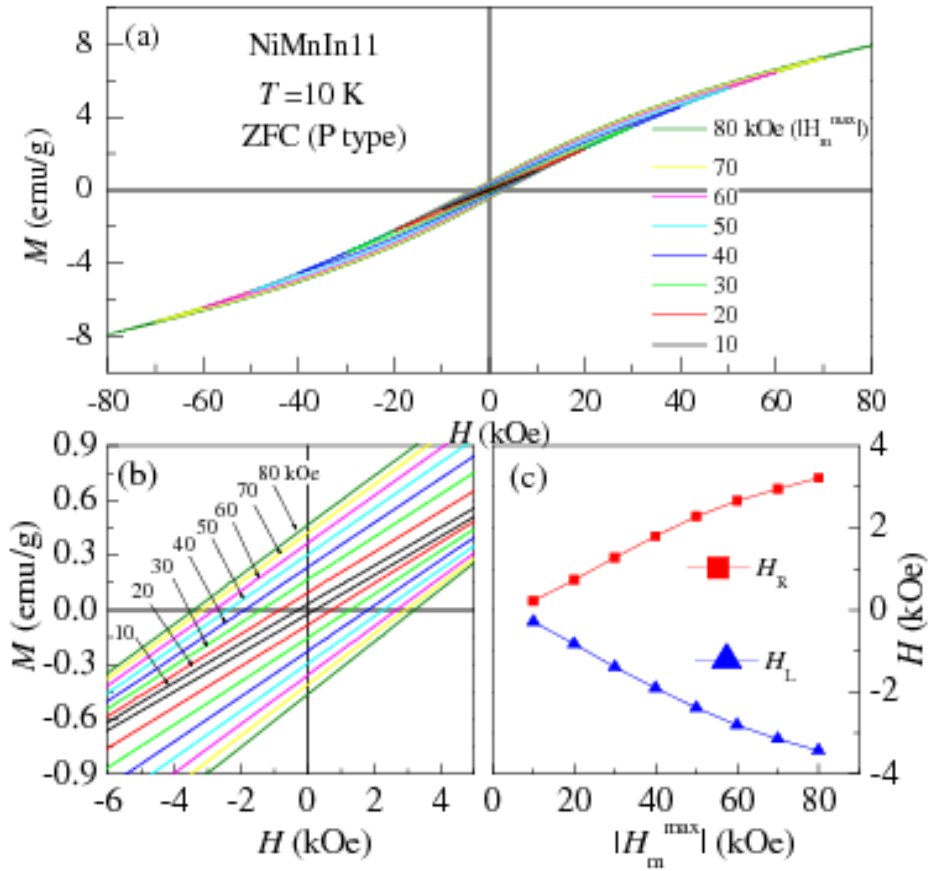

Fig. S27. (a) P type $M(H)$ loops measured under different $|H_m^{max}|$s in NiMnIn11 at 10 K after ZFC from 300 K. (b) The low field part of P type $M(H)$ curves for different $|H_m^{max}|$s. (c) The left ($H_L$) and right ($H_R$) coercive fields as a function of $|H_m^{max}|$.

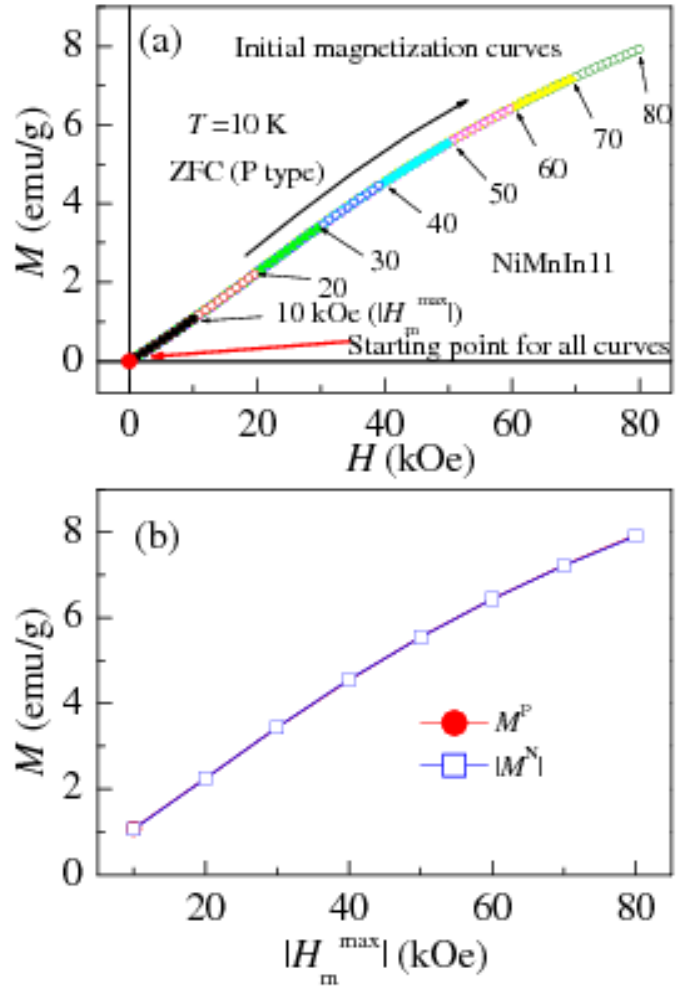

Fig. S28. (a) Initial magnetization curves [0 → (+$H$)] for different |$H_m^{max}$|s in NiMnIn11 at 10 K after ZFC from 300 K. (b) The magnetization values in the highest positive and negative magnetic fields for different |$H_m^{max}$|s.

**b. NiMnIn12**

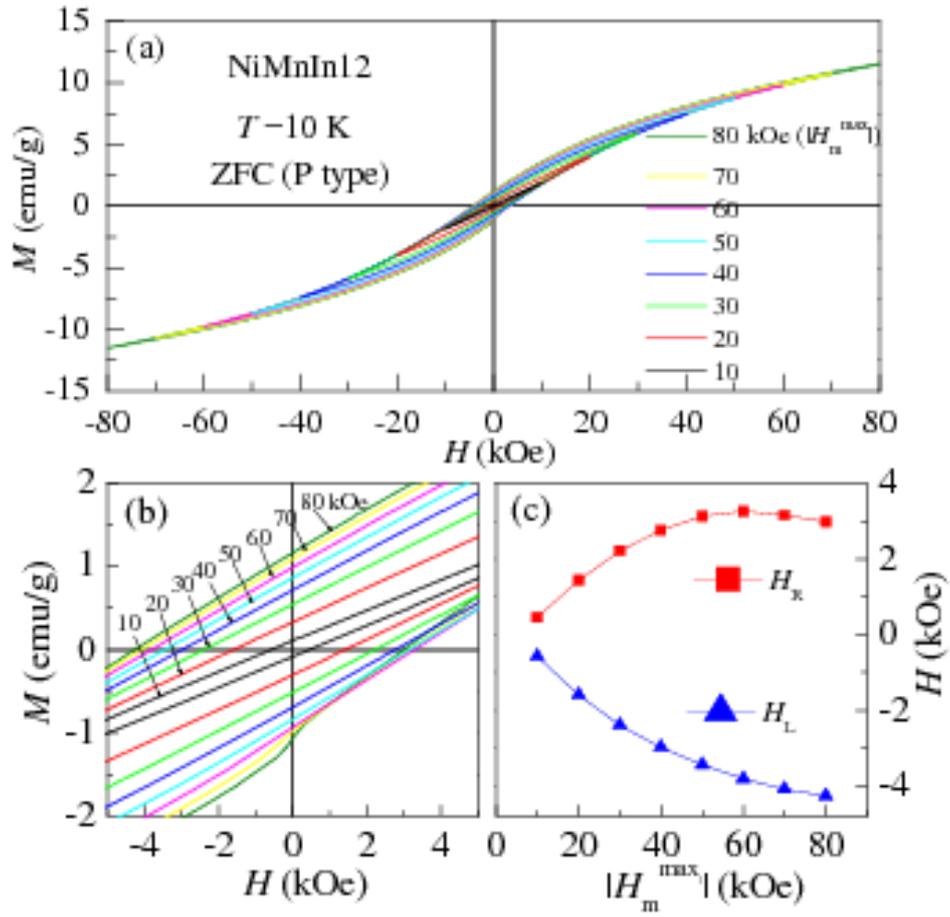

Fig. S29. (a) P type $M(H)$ loops measured under different $|H_m^{max}|$s in NiMnIn12 at 10 K after ZFC from 300 K. (b) The low field part of P type $M(H)$ curves for different $|H_m^{max}|$s. (c) The left ($H_L$) and right ($H_R$) coercive fields as a function of $|H_m^{max}|$.

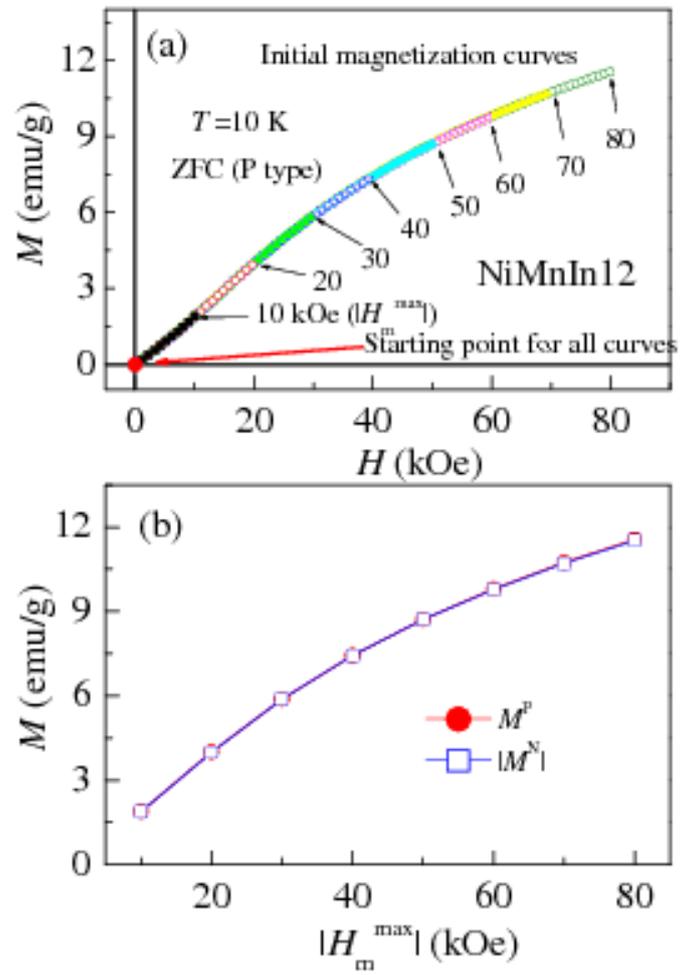

Fig. S30. (a) Initial magnetization curves [0 → (+$H$)] for different $|H_m^{max}|$s in NiMnIn12 at 10 K after ZFC from 300 K. (b) The magnetization values in the highest positive and negative magnetic fields for different $|H_m^{max}|$s.

## c. NiMnIn14

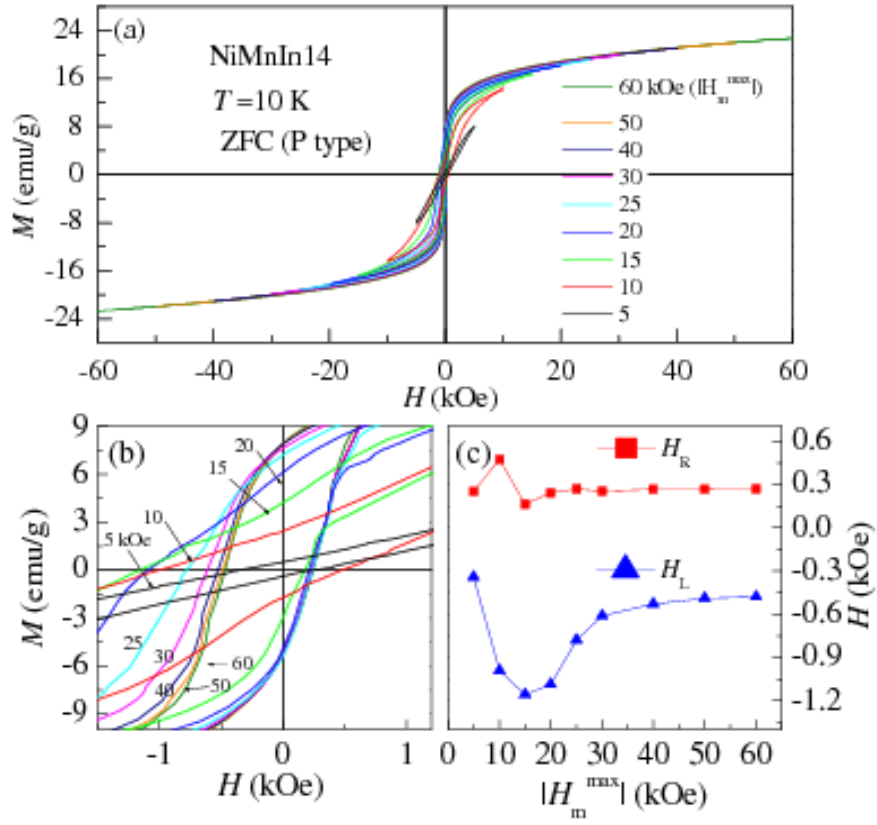

Fig. S31. (a) P type $M(H)$ loops measured under different $|H_m^{max}|$s in NiMnIn14 at 10 K after ZFC from 300 K. (b) The low field part of P type $M(H)$ curves for different $|H_m^{max}|$s. (c) The left ($H_L$) and right ($H_R$) coercive fields as a function of $|H_m^{max}|$.

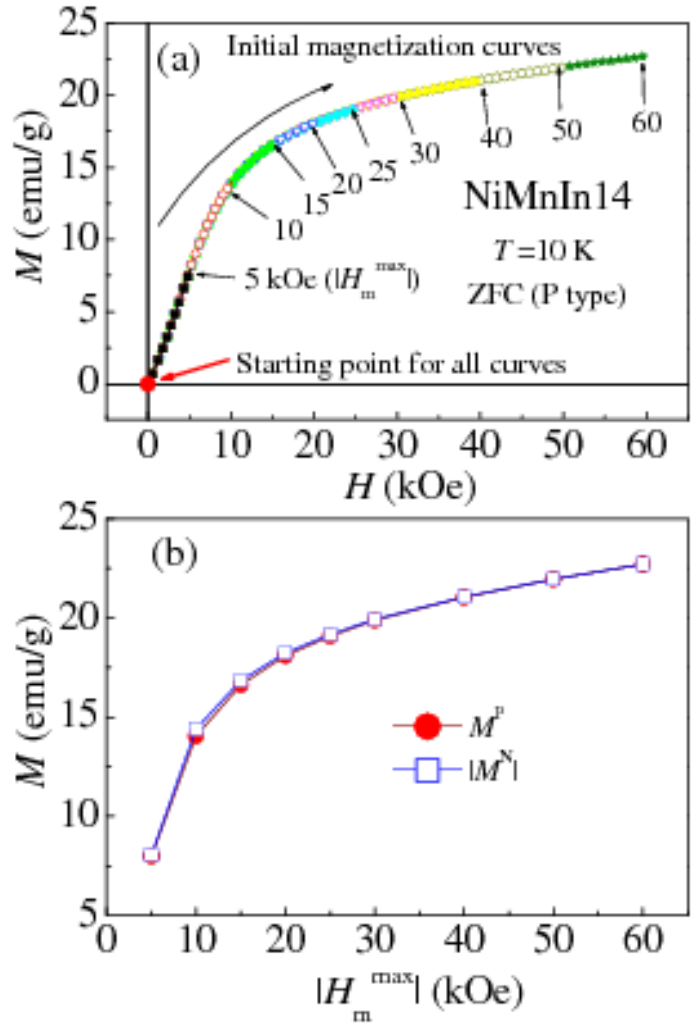

Fig. S32. (a) Initial magnetization curves [0 → (+H)] for different $|H_m^{max}|$s in NiMnIn14 at 10 K after ZFC from 300 K. (b) The magnetization values in the highest positive and negative magnetic fields for different $|H_m^{max}|$s.

### d. NiMnIn15

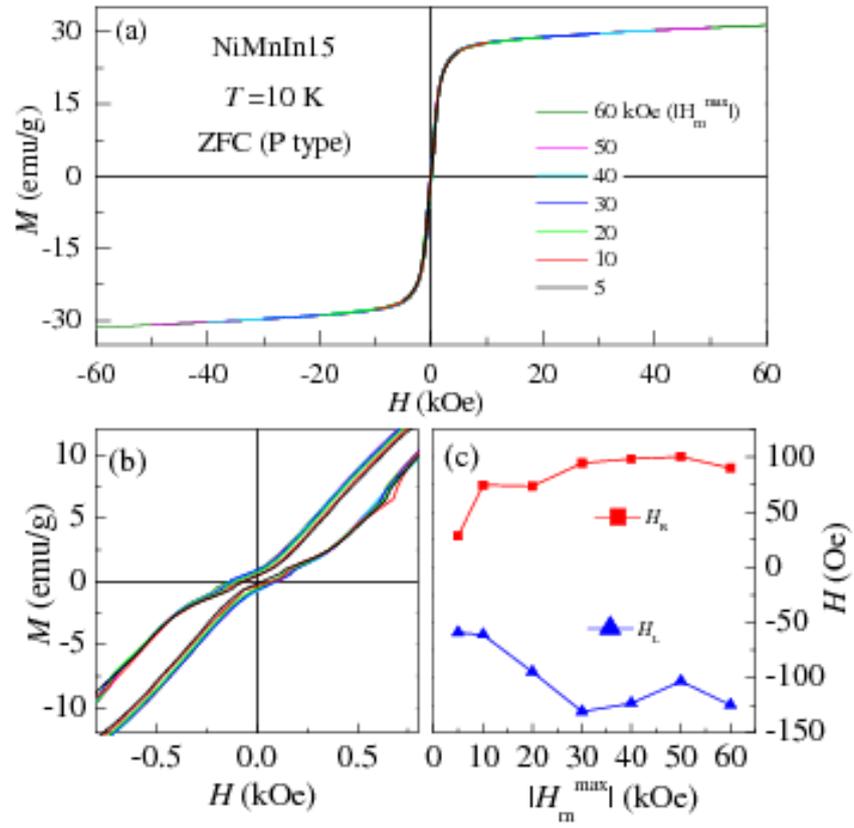

Fig. S33. (a) P type $M(H)$ loops measured under different $|H_m^{max}|$s in NiMnIn15 at 10 K after ZFC from 300 K. (b) The low field part of P type $M(H)$ curves for different $|H_m^{max}|$s. (c) The left ($H_L$) and right ($H_R$) coercive fields as a function of $|H_m^{max}|$.

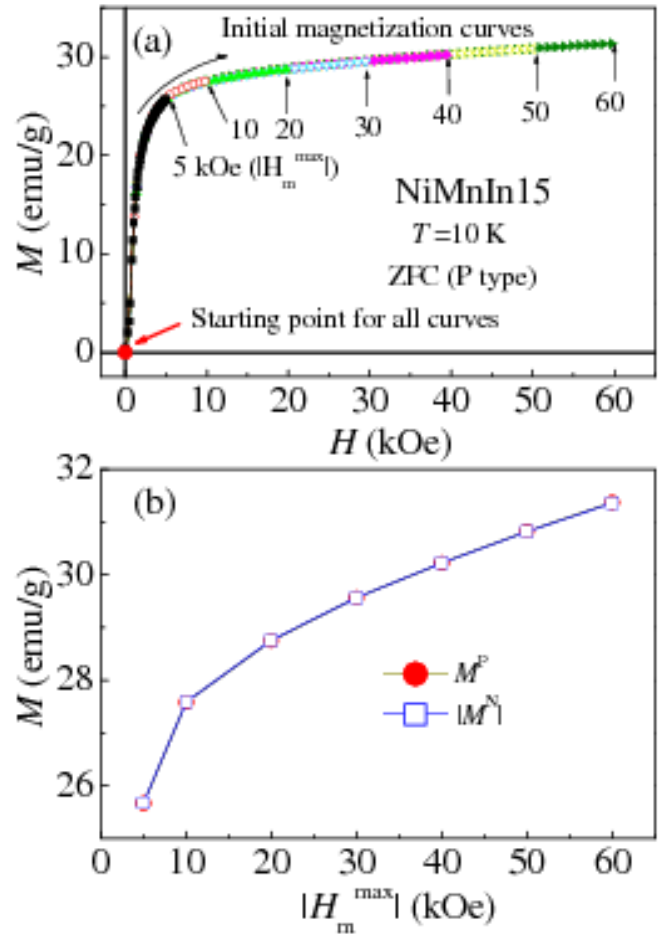

Fig. S34. (a) Initial magnetization curves [0 → (+H)] for different $|H_m^{max}|$s in NiMnIn15 at 10 K after ZFC from 300 K. (b) The magnetization values in the highest positive and negative magnetic fields for different $|H_m^{max}|$s.